\documentclass[11pt,a4paper]{article}
\pdfoutput=1
\usepackage{jcappub}

\usepackage{graphicx} 
\usepackage{caption}
\usepackage{subcaption}
\usepackage{slashed} 
\usepackage{amsfonts}
\usepackage{enumerate} 
\usepackage{color}
\usepackage{amsmath} 
\usepackage{bm}
\usepackage[paperwidth=210mm,paperheight=280mm,centering,hmargin=2.2cm,vmargin=2.2cm]{geometry}

\title{Non-linear curvature perturbation in multi-field inflation models with non-minimal coupling}

\author[a]{Jonathan White,}
\author[a,b]{Masato Minamitsuji}
\author[a]{and Misao Sasaki}

\affiliation[a]{Yukawa Institute for Theoretical Physics, Kyoto University\\Kyoto 606-8502, Japan}
\affiliation[b]{Multidisciplinary Center for Astrophysics (CENTRA), Instituto Superior T\'ecnico \\Lisbon 1049-001, Portugal.}

\emailAdd{jwhite@yukawa.kyoto-u.ac.jp}
\emailAdd{masato.minamitsuji@ist.utl.pt}
\emailAdd{misao@yukawa.kyoto-u.ac.jp}

 \abstract{Using the $\delta N$ formalism we consider the non-linear curvature
 perturbation in multi-field models of inflation with non-minimal
 coupling.  In particular, we focus on the relation between the $\delta
 N$ formalism as applied in the conformally related Jordan and Einstein
 frames.  Exploiting results already known in the Einstein frame, we
 give expressions for the power spectrum, spectral tilt and 
 non-gaussianity associated with the Jordan frame curvature perturbation.
 In the case that an adiabatic limit has not been reached, we find that
 in general these quantities differ from those associated with the
 Einstein frame curvature perturbation, and also confirm their
 equivalence in the absence of isocurvature modes.  We then proceed to
 consider two analytically soluble examples, the first involving a
 non-minimally coupled `spectator' field and the second being a
 non-minimally coupled extension of the multi-brid inflation model.  In
 the first model we find that predictions can easily be brought into
 agreement with the recent {\it Planck} results, as the tensor-to-scalar
 ratio is generally small, the spectral tilt tuneable and the non-gaussianity
 suppressed.  In the second model we find that predictions for all three
 parameters can differ substantially from those predicted in the
 minimally coupled case, and that the recent {\it Planck} results for
 the spectral tilt can be used to constrain the non-minimal coupling
 parameters.}
 
\arxivnumber{1306.6186}
\begin{document}
\maketitle

\section{Introduction}

An epoch of inflation is widely accepted as accounting for the physics
of the early Universe, with predictions from simple single-field
inflationary models still being consistent with observational data, including
the WMAP 9-year data \cite{wmap9,wmap92}
and the recent {\it Planck} results
\cite{%wmap9,wmap92,
planck,planck2,planck3}.  In the context of
unifying, high-energy theories, however, it is natural to expect
modifications to the simplest single-field scenarios in the form of multiple fields,
non-canonical kinetic terms or modifications to the gravity sector.  It
is thus important to determine the observational signatures of such
modifying features in order that we may constrain different
models of inflation and the high-energy theories that motivate them.

One particular observable that can potentially be used to distinguish
between different models of inflation is the non-gaussianity of the
Cosmic Microwave Background, which is closely related to the
non-gaussianity of the curvature perturbation on constant energy
hypersurfaces, $\zeta$.  It is known that the non-gaussianity predicted
by the simplest single-field models is slow-roll suppressed \cite{mal},
so its non-negative detection would immediately point to something
beyond the simplest models.  Moreover, the shape and size of
non-gaussianity predicted by more complicated models depend on the types
of modifications considered, thus deeming it a very useful discriminant.
The recent {\it Planck} results, which are still consistent with
gaussian fluctuations, put tight constraints on inflationary models
predicting large non-gaussianity, but still allow for deviations from
the simplest of single-field scenarios \cite{planck,planck2,planck3,suyama}.

In this paper we are interested in the non-linear curvature perturbation
in multi-field models of inflation with non-minimal coupling to the
gravity sector.  The type of models that we consider take an action of
the form
\begin{equation}\label{JAc}
S = \int d^4x\sqrt{-g}\left\{f({\bm \phi})R - \frac{1}{2}h_{ab}({\bm
\phi})g^{\mu\nu}\partial_\mu\phi^a\partial_\nu\phi^b - V({\bm
\phi})\right\},
\end{equation}
where $R$ is the scalar curvature associated with the metric $g_{\mu\nu}$, $a, b = 1, ... , n$ 
label $n$ scalar fields that are potentially
all non-minimally coupled to gravity through $f(\bm{\phi})$ and the
non-diagonal field-space metric $h_{ab}$ gives a non-canonical kinetic
term.  Such a form of action is well motivated in the context of
higher-dimensional theories, where non-minimal coupling and
non-canonical kinetic terms appear naturally by way of compactification
\cite{cew}.  Non-minimal couplings also arise in the context of
renormalisation \cite{ccj,fm}.

The action in its original form \eqref{JAc} is said to be in the Jordan
frame, but by making the conformal transformation $g_{\mu\nu} =
(2f)^{-1}\tilde{g}_{\mu\nu}$ we are able to move to the Einstein frame,
where the action takes the form
\begin{equation}\label{EAc}
S = \int d^4x\sqrt{-\tilde{g}}\left\{\frac{\tilde{R}}{2}
-\frac{1}{2}S_{ab}(\bm{\phi})\tilde{g}^{\mu\nu}\partial_\mu\phi^a\partial_\nu\phi^b
- \tilde{V}(\bm{\phi})\right\},
\end{equation}     
where $\tilde{R}$ is the scalar curvature associated with the  
Einstein frame metric $\tilde{g}_{\mu\nu}$,\footnote{Note that as the fields are minimally coupled in the Einstein frame, it is the metric $S_{ab}$ rather than $h_{ab}$ that is important in determining the ghost-free condition. 
We thus require $S_{ab}$ to be positive-definite with Euclidean signature
(see also footnote 5 of our previous paper \cite{wms}).
}   
\begin{equation}
S_{ab} = \frac{1}{2f}\left(h_{ab} +
\frac{3f_af_b}{f}\right)\qquad\mbox{and}\qquad \tilde{V}(\bm{\phi})=
\frac{V}{4f^2}.
\label{albert}
\end{equation} 
Here $f_a$ denotes the derivative of $f$ with respect to the $a$'th
field and we have taken $\kappa^2 =8\pi G= 1$.  The action \eqref{EAc}, with
its canonical gravity sector, is one we are more familiar with, and
computations are seemingly more tractable in this frame.  Indeed, a
framework for the calculation of $\tilde{\zeta}$ and its non-gaussianity
for an action of the form \eqref{EAc} has recently been developed in
\cite{gong2, eliston,Langlois} and applied to some well motivated examples in \cite{kaiser,kaiser2}.  
%It was
%found that 
%The form of potential $\tilde{V}$ induced by the conformal
%transformation naturally displays ridge-like features, the presence of
%which is known to allow for the possibility of large non-gaussianities
%in multi-field inflation models 
The effective potential $\tilde V$ induced by the conformal
transformation can display interesting structures, including ridge-like
features, which are known to potentially give rise to large
non-gaussianity \cite{kaiser}.  Furthermore, we see that even if the
Jordan frame field-space metric is taken to be flat, i.e. $h_{ab} =
\delta_{ab}$, the induced Einstein frame field-space is not flat.  This
leads to additional couplings between the fields that can also
potentially source non-gaussianity.  With a framework for performing
calculations in the Einstein frame already in place, in this paper we
are interested in how the quantity $\tilde{\zeta}$ can be related back
to $\zeta$ in the original Jordan frame.  In the single-field case it is
known that $\zeta = \tilde{\zeta}$ to all orders of perturbation
\cite{ms,gong,cy}, but in the multi-field case this is no longer true,
with the difference being a direct consequence of the isocurvature modes
inherent to multi-field inflation \cite{gordon,wms}.  In the earlier
paper \cite{wms} we have discussed the relationship between $\zeta$ and
$\tilde{\zeta}$ at linear order, highlighting that both the evolution
and magnitude of the two quantities can potentially be very different.
We argued that if isocurvature modes are still present at the epoch of
last scattering, the difference between $\zeta$ and $\tilde{\zeta}$
simply highlights their non-observable nature, as the predictions for
any observable quantity should be independent of the frame in which they
are calculated, despite any possible differences in physical
interpretation \cite{ndms,catena}.  It is therefore important to take into
account to which metric matter is minimally coupled and how $\zeta$ or
$\tilde{\zeta}$ at the epoch of last scattering are related to what we
actually observe today.  In this paper we extend the analysis of
\cite{wms} beyond linear order.

The method that we adopt is the $\delta N$ formalism \cite{starobinsky,ss,ns,st,lms}, which states that
on super-horizon scales the %non-perturbative
nonlinear curvature perturbation
$\zeta$ is given as 
\begin{equation}
\zeta(t_\diamond, {\bm x})= \delta N = N(t_\ast, t_\diamond;{\bm x}) -
N_0(t_\ast, t_\diamond),
\end{equation}
where $N_0(t_\ast, t_\diamond)$ is the background number of e-foldings
between the initial time $t_\ast$ and final time $t_\diamond$ and
$N(t_\ast, t_\diamond;{\bm x})$ the number of e-foldings between a flat
hypersurface at the initial time $t_\ast$ and a constant energy
hypersurface at the final time $t_\diamond$.
The time $t_\ast$ corresponds to a time shortly after the scales under
consideration have left the horizon.  Under the separate universe
approximation, only the background equations of motion are required to
determine the local number of e-foldings for each patch, and the
difference in e-folding number between different patches is determined
by the difference in initial conditions, which are in turn determined by
the requirement that the initial hypersurface be flat.  In the case of
multi-field inflation we can write $N = N(\bm{\phi})$, and thus $\delta
N$ is determined by the values of the fields $\phi^a$ on the initial
flat slice.  Expanding perturbatively we have
\begin{equation}\label{JFexpan}
\zeta = \delta N = N_a\delta\phi^a_{\mathcal{R}} +
\frac{1}{2}N_{ab}\delta\phi^a_{\mathcal{R}}\delta\phi^b_{\mathcal{R}} \,\, ...\,,
\end{equation}
where $N_a$ indicates the partial derivative of $N$ with respect to the
initial value of the $a$'th field and the subscript $\mathcal{R}$ indicates the
field perturbations on the initial flat slice.  Note that here we have
made the slow-roll approximation such that the initial field velocities are not 
considered as independent degrees of freedom, i.e $\dot{\phi}^{a} =
\dot{\phi}^{a}(\bm{\phi})$.

If expression \eqref{JFexpan} is for the curvature perturbation in the
Jordan frame, then in the Einstein frame we have the similar expression
\begin{equation}\label{EFexpan}
\tilde{\zeta} = \delta \tilde{N} = \tilde{N}_a\delta\phi^a_{\tilde{\mathcal{R}}} +
\frac{1}{2}\tilde{N}_{ab}\delta\phi^a_{\tilde{\mathcal{R}}}\delta\phi^b_{\tilde{\mathcal{R}}}
\,\, ...\,,
\end{equation}
where $N\rightarrow\tilde{N}$ and $\delta\phi^a_{\mathcal{R}}\rightarrow
\delta\phi^a_{\tilde{\mathcal{R}}}$.  There therefore appear to be two
sources for discrepancies between the two curvature perturbations:
differences in the derivatives of $N$ and $\tilde{N}$ with respect to
the initial conditions and differences in the definition of the initial
flat slice on which the field perturbations are evaluated.  In this
paper, by considering the definition of gauge-independent variables up
to second order, we first determine the relation between
$\delta\phi^a_{\mathcal{R}}$ and $\delta\phi^a_{\tilde{\mathcal{R}}}$.
Then, by considering the definitions of $N$ and $\tilde{N}$ as the
integrals of $H$ and $\tilde{H}$ respectively, we are also able to
determine the relation between the derivatives of $N$ and $\tilde{N}$
with respect to the initial field values.  What we find is that the
effect of the difference in definition of the initial flat hypersurfaces
exactly cancels with one of the terms arising from the difference in
definition of $N$ and $\tilde{N}$.  As such, the only remaining
difference between $\zeta$ and $\tilde{\zeta}$ is that due to the
difference in definition of the final constant energy surface in
the two frames.  In the adiabatic limit, where the definition of the
final constant energy surface coincides, the expected result $\zeta
= \tilde{\zeta}$ is therefore recovered.

Turning to the calculation of the correlation functions of $\zeta$ and
$\tilde{\zeta}$, we see that
given the above expansions \eqref{JFexpan} and \eqref{EFexpan} in terms
of the field perturbations on the initial flat slices, the correlation
functions of $\zeta$ or $\tilde{\zeta}$ can be calculated if we know the
correlation functions of the field perturbations.  The two-point
correlation function of $\delta\phi^a_{\tilde{\mathcal{R}}}$ has been known at
linear order for some time \cite{ss,ns}, 
and recently Elliston et al. \cite{eliston}
have also determined the
three-point correlation function of $\delta\phi^a_{\tilde{\mathcal{R}}}$
for an action of the form \eqref{EAc} (see also \cite{Langlois}).
As such, exploiting the relation we find between $\delta\phi^a_{\mathcal{R}}$ and
$\delta\phi^a_{\tilde{\mathcal{R}}}$, we are able to give general expressions
for the two- and three-point correlation functions of $\zeta$ as well as
$\tilde{\zeta}$. 

After determining general expressions for $\zeta$ and $\tilde{\zeta}$
and their associated power spectra, spectral tilts and $f_{NL}$
parameters, we then focus on some analytically soluble examples that
allow us to directly compute and compare $\zeta$ and $\tilde{\zeta}$.
The first example we consider is a two-field model similar to that
considered at linear order in \cite{wms}, where we investigated the
possibility that the curvature perturbation in the Jordan and Einstein
frames have very different evolutions.  In addition to the minimally
coupled inflaton field $\phi$, the model consists of a non-minimally
coupled spectator field $\chi$, where by `spectator' field we mean that
the field is non-dynamical during inflation.  The fact that the $\chi$
field is non-minimally coupled, however, means that despite being non-dynamical at 
background level it does affect the curvature
perturbation throughout inflation.  As such, $\chi$ is not a spectator field in the 
usual sense.  We find that the
curvature perturbation in the Jordan and Einstein frames do indeed
evolve very differently, as do the corresponding power spectra, spectral
tilts and $f_{NL}$ parameters, but that by the end of inflation the two
quantities converge to leading order in the slow-roll approximation.
Regarding the final predictions for the curvature perturbation and its
spectral properties, we find that they can easily be brought into agreement with
the recent {\it Planck} results.  In particular, the presence of the
$\chi$ field tends to act so as to reduce the tensor-to-scalar ratio, as
well as allowing for a wide range of spectral tilts.  In the case that
the Einstein frame field-space curvature is assumed to be small, the model
predicts a very small non-gaussianity.

The second example we consider is a non-minimally coupled extension of
the multi-brid inflation model introduced in \cite{multibrid}.  In this
model, because the end-of-inflation surface is defined by the
tachyonic instability of a waterfall field instead of a constant energy surface, we find that the curvature
perturbation in the Jordan and Einstein frames are identical.  Choosing
the same model parameters as the original model, we find that the
introduction of non-minimal coupling has a significant effect on the
predicted power spectrum, spectral tilt and non-gaussianity.  Whilst the tensor-to-scalar ratio and
$f_{NL}$ parameter remain in agreement with the recent {\it Planck}
results, requiring agreement with the observed spectral tilt places constraints on the form of
non-minimal coupling.

The paper is constructed as follows: In section \ref{generalformalism}, applying the $\delta N$
formalism in both the Jordan and Einstein frames we obtain general
expressions for the curvature perturbation in each frame and compare the
results.  In section \ref{analsec} we discuss the class of analytically soluble
models where both the potential and non-minimal coupling function are of
product separable form.  In section \ref{nmcssec} and section \ref{nmmbsec} we focus on the two explicit examples
mentioned above, and in section \ref{concsec} we close with a brief summary and
conclusions.

\section{Applying the $\delta N$ formalism in the Jordan and Einstein frames}\label{generalformalism}

In this section we consider the relation between the two key elements of
the $\delta N$ formalism as applied in the Jordan and Einstein frames.
Namely, we would like to know how derivatives of $N$ with respect to the
initial conditions are related those of $\tilde{N}$ and also how the
flat-gauge field perturbations of the Jordan frame are related to those
in the Einstein frame.

\subsection{Curvature perturbation in the Jordan frame}

Let us begin by discussing the Jordan frame.  
Following \cite{kodamasasaki}, we define our perturbed Jordan frame
metric as
\begin{equation}\label{metDecomp}
%\frac{ds^2}{a^2}
ds^2 =
a(\eta)^2
\left\{ -(1+2AY)d\eta^2 -2BY_i d\eta dx^i
+\left[\left(1+2\mathcal{R}\right)\delta_{ij}+2H_T\frac{1}{k^2}Y_{,ij}\right]dx^idx^j\right\},
\end{equation}
where $d\eta = dt/a$ is the conformal time,
$a$ and $t$ are the scale factor and the proper time in the Jordan frame, respectively,
%we use conformal time defined as
the scalar harmonic 
functions $Y$ satisfy $(\nabla^2+k^2)Y=0$ and $Y_i = -k^{-1}Y_{,i}$ \cite{kodamasasaki},
and each of the perturbation quantities is
decomposed in terms of first- and second-order components as
\begin{equation}
\mathcal{R} = \mathcal{R}_{(1)} + \frac{1}{2}\mathcal{R}_{(2)}.
\end{equation}
Note that here we neglect vector and tensor modes  and only consider scalar perturbations.  We expect this simplifying assumption to be valid on super-horizon scales.  On super-horizon
scales, by considering gauge transformations up to second order, we find
the Sasaki-Mukhanov variables \cite{sasakimukhanov1,sasakimukhanov2}, or
flat-gauge field perturbations, to be given as \cite{Malik}
\begin{align}\label{JFflatperts}
\delta\phi_{\mathcal{R}(1)}^a &= \delta\phi^a_{(1)} - \frac{\phi^{\prime
a}}{\mathcal{H}}\mathcal{R}_{(1)},\\\nonumber \delta\phi_{\mathcal{R}(2)}^a& =
\delta\phi^a_{(2)} -\frac{\phi^{\prime a}}{\mathcal{H}}\mathcal{R}_{(2)}
+
\left(\frac{\mathcal{R}_{(1)}}{\mathcal{H}}\right)^2\left[2\mathcal{H}\phi^{\prime
a} + \phi^{\prime\prime a} -
\frac{\mathcal{H}^\prime}{\mathcal{H}}\phi^{\prime a}\right] +
2\frac{\phi^{\prime
a}}{\mathcal{H}^2}\mathcal{R}^\prime_{(1)}\mathcal{R}_{(1)}-
\frac{2}{\mathcal{H}}\mathcal{R}_{(1)}\delta\phi^{\prime a}_{(1)},
\end{align}
where %we are using
a prime  denotes the derivative with respect to conformal time, $\mathcal{H} =
a^\prime/a$ and the subscript $\mathcal{R}$ denotes the flat-gauge field
perturbations in the Jordan frame.

Similarly, we define the perturbed Einstein frame metric as
\begin{equation}\label{metDecompE}
%\frac{d\tilde{s}^2}{\tilde{a}^2}
d\tilde{s}^2 =
\tilde{a}(\eta)^2
\left\{ -(1+2\tilde{A}Y)d\eta^2 -2\tilde{B}Y_i d\eta dx^i
+\left[\left(1+2\tilde{\mathcal{R}}\right)\delta_{ij}+2\tilde{H}_T\frac{1}{k^2}Y_{,ij}\right]dx^idx^j
\right\},
\end{equation}
where $\tilde{a}=\sqrt{2f}a$ and $d\tilde{t}=\sqrt{2f}dt$ are the scale factor and the proper time in the Einstein frame,
respectively,
$d\eta = d\tilde{t}/\tilde{a}=dt/a$ is again the conformal time, 
and each of the perturbation quantities is again
decomposed in terms of first- and second-order components as
\begin{equation}
\tilde{\mathcal{R}} =
\tilde{ \mathcal{R}}_{(1)} 
+ \frac{1}{2}\tilde{\mathcal{R}}_{(2)}.
\end{equation}
The flat-gauge field perturbations in the Einstein frame, 
$\delta\phi_{\tilde{\mathcal{R}}(1)}^a$ and $\delta\phi_{\tilde{\mathcal{R}}(2)}^a$,
are then found by replacing all $\mathcal{R}$ and $\mathcal{H}$ in \eqref{JFflatperts}
 with $\tilde{\mathcal{R}}$ and $\tilde{\mathcal{H}}=\tilde{a}^{\prime}/\tilde{a}$
 respectively. 

In order to find the relation between the flat-gauge field perturbations
in the Jordan and Einstein frames, we first need to determine the
relation between $\mathcal{R}$ and $\tilde{\mathcal{R}}$ to second
order.  Using the fact that $d\tilde{s}^2 = 2f ds^2$ we find that
\begin{align}
\tilde{\mathcal{R}}_{(1)} &= \mathcal{R}_{(1)} +
\frac{f_{a}\delta\phi^a_{(1)}}{2f}\\
\mbox{and}\qquad\tilde{\mathcal{R}}_{(2)} &= \mathcal{R}_{(2)} +
\frac{f_{a}\delta\phi^a_{(2)}}{2f}+\frac{f_{ab}\delta\phi^a_{(1)}\delta\phi^b_{(1)}}{2f}-\frac{f_{a}f_{b}\delta\phi^a_{(1)}\delta\phi^b_{(1)}}{f^2}
+ \frac{2f_a\delta\phi^a_{(1)}\tilde{\mathcal{R}}_{(1)}}{f}.
\end{align}  
Substituting these relations, along with the background result $\tilde{\mathcal{H}} =
\mathcal{H} + \frac{f^\prime}{2f}$, into \eqref{JFflatperts} and evaluating the
right-hand side of each equation in the flat gauge of the Einstein frame (recall
that we can evaluate the right-hand side of the equations in whichever gauge we
like, as overall the expressions are gauge-invariant) we
deduce
%\footnote{One can explicitly check these results at linear order
%by substituting \eqref{JtoE1} into the first-order equations of motion
%derived in \cite{ns} for the $\delta\phi^a_{f(1)}$ and seeing that this
%reproduces the corresponding equations of motion for
%$\delta\phi^a_{\tilde{\mathcal{R}}(1)}$ derived in the Einstein frame. 
%}
\begin{align}\label{JtoE1}
\delta\phi^a_{\mathcal{R}(1)} &= \Pi^a{}_b\delta\phi^b_{\tilde{\mathcal{R}}(1)},\\\nonumber
\delta\phi^a_{\mathcal{R}(2)} &= \Pi^a{}_b\delta\phi^b_{\tilde{\mathcal{R}}(2)} +
\frac{\phi^{\prime
a}}{\mathcal{H}}\frac{f_{cb}\delta\phi^c_{\tilde{\mathcal{R}}(1)}\delta\phi^b_{\tilde{\mathcal{R}}(1)}}{2f}
- \frac{2\phi^{\prime
a}}{\mathcal{H}}\left(\frac{f_b\delta\phi^b_{\tilde{\mathcal{R}}(1)}}{2f}\right)^2
+ \frac{1}{\mathcal{H}}\frac{d}{d\eta}\left[\frac{\phi^{\prime
a}}{\mathcal{H}}\left(\frac{f_b\delta\phi^b_{\tilde{\mathcal{R}}(1)}}{2f}\right)^2\right]\\\label{JtoE2}
&\hspace{2.4cm}+\frac{2}{\mathcal{H}}\frac{f_b\delta\phi^b_{\tilde{\mathcal{R}}(1)}\delta\phi^{\prime
a}_{\tilde{\mathcal{R}}(1)}}{2f},
\end{align}
where $\Pi^a{}_b = \delta^a_b + \frac{\phi^{\prime a}
f_b}{2f\mathcal{H}}$ and the subscript $\tilde{\mathcal{R}}$ denotes the
flat-gauge field perturbations in the Einstein frame.

On inserting \eqref{JtoE1} and \eqref{JtoE2} into the expansion to
second order of $\delta N \, (=\zeta)$, we find
\begin{align}
\delta N &= N_a\Pi^a{}_b\delta\phi^b_{\tilde{\mathcal{R}}} +
\frac{1}{2}N_{ab}\Pi^a{}_c\Pi^b{}_d\delta\phi^c_{\tilde{\mathcal{R}}(1)}\delta\phi^d_{\tilde{\mathcal{R}}(1)}\\\nonumber
&\quad + \frac{N_a}{2}\left(\frac{\phi^{\prime
a}}{\mathcal{H}}\frac{f_{cb}\delta\phi^c_{\tilde{\mathcal{R}}(1)}\delta\phi^b_{\tilde{\mathcal{R}}(1)}}{2f}
- \frac{2\phi^{\prime
a}}{\mathcal{H}}\left(\frac{f_b\delta\phi^b_{\tilde{\mathcal{R}}(1)}}{2f}\right)^2
\right.\\\nonumber&\quad\left.+ \frac{1}{\mathcal{H}}\frac{d}{d\eta}\left[\frac{\phi^{\prime
a}}{\mathcal{H}}\left(\frac{f_b\delta\phi^b_{\tilde{\mathcal{R}}(1)}}{2f}\right)^2\right]+\frac{2}{\mathcal{H}}\frac{f_b\delta\phi^b_{\tilde{\mathcal{R}}(1)}\delta\phi^{\prime
a}_{\tilde{\mathcal{R}}(1)}}{2f}\right),
%&=N_a\delta\phi^a_{\tilde{f}} + \frac{1}{2}N_{ab}\delta\phi^a_{(1)\tilde{f}}\delta\phi^b_{(1)\tilde{f}}+\frac{f_a\delta\phi^a_{\tilde{f}}}{2f} + \frac{1}{2}\frac{f_{cb}\delta\phi^c_{\tilde{f}(1)}\delta\phi^b_{\tilde{f}(1)}}{2f}- \left(\frac{f_b\delta\phi^b_{\tilde{f}(1)}}{2f}\right)^2\\
%&\quad + N_{ab}\frac{\phi^{\prime a}f_c}{2f\mathcal{H}}\delta\phi^c_{\tilde{f}(1)}\delta\phi^b_{\tilde{f}(1)} + \frac{1}{2}N_{ad}\frac{\phi^{\prime a}\phi^{\prime d}}{4f^2\mathcal{H}^2}f_cf_b\delta\phi^c_{\tilde{f}(1)}\delta\phi^b_{\tilde{f}(1)}\\&\quad+ \frac{N_a}{2\mathcal{H}}\frac{d}{d\eta}\left[\frac{\phi^{\prime a}}{\mathcal{H}}\left(\frac{f_b\delta\phi^b_{\tilde{f}(1)}}{2f}\right)^2\right]+\frac{N_a}{\mathcal{H}}\frac{f_b\delta\phi^b_{\tilde{f}(1)}\delta\phi^{\prime a}_{\tilde{f}(1)}}{2f},
\end{align}
which, using $N_a\phi^{\prime a} = dN/d\eta_\ast = -\mathcal{H}$ (see \eqref{N}), can be
simplified to
\begin{align}\nonumber
\delta N&=N_a\delta\phi^a_{\tilde{\mathcal{R}}} +
\frac{1}{2}N_{ab}\delta\phi^a_{\tilde{\mathcal{R}}(1)}\delta\phi^b_{\tilde{\mathcal{R}}(1)}-\frac{f_a\delta\phi^a_{\tilde{\mathcal{R}}}}{2f}
-
\frac{1}{2}\frac{f_{cb}\delta\phi^c_{\tilde{\mathcal{R}}(1)}\delta\phi^b_{\tilde{\mathcal{R}}(1)}}{2f}+
\left(\frac{f_b\delta\phi^b_{\tilde{\mathcal{R}}(1)}}{2f}\right)^2\\ &\quad +
\frac{1}{\mathcal{H}}\frac{f_a\delta\phi^a_{\tilde{\mathcal{R}}(1)}}{2f}\frac{d}{d\eta_\ast}\left[\delta
 N_{(1)}\right].
\end{align}
Note that, except for $N_a$ and $N_{ab}$, which in general will depend 
on the final time as well as the initial time $\eta_\ast$, all quantities in
the above expansion are to be evaluated at an initial time $\eta_\ast$
shortly after the modes in question have left the horizon.
%In our working so far, we have implicitly made the slow-roll assumption
%by only expanding $N$ in terms of the initial $\phi^a$.  As such, we can
However, through the $\delta N$ formalism we know that 
%it can be shown that
$\delta N_{(1)} = \zeta_{(1)}(\eta_\diamond,x^i)$, where $\eta_\diamond$ denotes the final time, 
i.e. $\delta N_{(1)}$ does not depend on the initial time $\eta_\ast$.\footnote{This independence of $\eta_\ast$ has been shown explicitly up to second order in the slow-roll approximation in \cite{ns}.} 
As such, the term on the second line of the above
expression vanishes, giving us
\begin{align}\label{Jzeta}
\zeta = \delta N=N_a\delta\phi^a_{\tilde{\mathcal{R}}} +
\frac{1}{2}N_{ab}\delta\phi^a_{\tilde{\mathcal{R}}(1)}\delta\phi^b_{\tilde{\mathcal{R}}(1)}-\frac{f_a\delta\phi^a_{\tilde{\mathcal{R}}}}{2f}
-
\frac{1}{2}\frac{f_{cb}\delta\phi^c_{\tilde{\mathcal{R}}(1)}\delta\phi^b_{\tilde{\mathcal{R}}(1)}}{2f}+\left(\frac{f_b\delta\phi^b_{\tilde{\mathcal{R}}(1)}}{2f}\right)^2.
\end{align}

Next, in order to allow us to use the results of Elliston et al. \cite{eliston}
to
calculate correlation functions, we need to use the relationship between
the $\delta\phi^a_{\tilde{\mathcal{R}}}$ used above and the corresponding
perturbation quantities that transform as vectors of the tangent space
defined by the field-space metric $S_{ab}$.  The relation to second
order is \cite{gong2}
\begin{equation}
\delta\phi^a_{\tilde{\mathcal{R}}} = Q^a_{\tilde{\mathcal{R}}} - \frac{1}{2}
{}^{(S)}\Gamma^a_{bc}Q^b_{\tilde{\mathcal{R}}(1)}Q^c_{\tilde{\mathcal{R}}(1)},
\end{equation} 
where ${}^{(S)}\Gamma^a_{bc}$ is the Christoffel symbol associated with $S_{ab}$ (we use ${}^{(h)}\Gamma^a_{bc}$ to denote that associated with $h_{ab}$), meaning that we now have
\begin{align}\nonumber
\zeta &= N_a\left(Q^a_{\tilde{\mathcal{R}}} - \frac{1}{2}
 {}^{(S)}\Gamma^a_{bc}Q^b_{\tilde{\mathcal{R}}(1)}Q^c_{\tilde{\mathcal{R}}(1)}\right) +
 \frac{1}{2}N_{ab}Q^a_{\tilde{\mathcal{R}}}Q^b_{\tilde{\mathcal{R}}}-\frac{f_a}{2f}\left(Q^a_{\tilde{\mathcal{R}}}
 - \frac{1}{2}
 {}^{(S)}\Gamma^a_{bc}Q^b_{\tilde{\mathcal{R}}(1)}Q^c_{\tilde{\mathcal{R}}(1)}\right)\\\nonumber
 &\quad-
 \frac{1}{2}\frac{f_{cb}Q^c_{\tilde{\mathcal{R}}(1)}Q^b_{\tilde{\mathcal{R}}(1)}}{2f}+\left(\frac{f_bQ^b_{\tilde{\mathcal{R}}(1)}}{2f}\right)^2\\
 &=\left(N_a - \frac{f_a}{2f}\right)Q^a_{\tilde{\mathcal{R}}} +
 \frac{1}{2}\left(\tilde{\nabla}_a\tilde{\nabla}_bN - \frac{\tilde{\nabla}_a\tilde{\nabla}_bf}{2f} +
 \frac{f_af_b}{2f^2}\right)Q^a_{\tilde{\mathcal{R}}(1)}Q^b_{\tilde{\mathcal{R}}(1)}.
\end{align}
Note that the covariant derivatives here are with respect to the
Einstein frame field-space metric.  Defining $\mathcal{N}=N-\ln(f)/2$, such that
\begin{equation}
\mathcal{N}_a = N_a - \frac{f_a}{2f}\qquad\mbox{and}\qquad
\tilde{\nabla}_a\tilde{\nabla}_b\mathcal{N} = \tilde{\nabla}_a\tilde{\nabla}_bN - \frac{\tilde{\nabla}_a\tilde{\nabla}_bf}{2f} +
\frac{f_af_b}{2f^2},
\end{equation}
and using the result \cite{ss,ns,eliston}
\begin{equation}
\langle Q^a_{\tilde{\mathcal{R}}(1)}(\bm{k}_1)Q^b_{\tilde{\mathcal{R}}(1)}(\bm{k}_2)\rangle
= (2\pi)^3\delta^3({\bm k}_1 + {\bm k}_2)\frac{\tilde{H}^2}{2k^3}S^{ab},
\end{equation}
we find the power spectrum and non-gaussianity in the Jordan frame
\begin{equation}\label{eq:1}
\mathcal{P}_\zeta(k) =
\mathcal{N}_a\mathcal{N}_bS^{ab}\left(\frac{\tilde{H}}{2\pi}\right)^2\qquad\mbox{and}\qquad
f_{NL} =\frac{5}{6}
\frac{\mathcal{N}_a\mathcal{N}_b\tilde{\nabla}_c\tilde{\nabla}_d\mathcal{N}S^{ac}S^{bd}}{\left[\mathcal{N}_e\mathcal{N}_{f}S^{ef}\right]^2},
\end{equation}
where, at leading order, the three point correlation function for
$\zeta$ is given as
\begin{align}
\langle\zeta(\bm{k}_1)\zeta(\bm{k}_2)\zeta(\bm{k}_3)\rangle &=
\mathcal{N}_a\mathcal{N}_b\mathcal{N}_c\langle\mathcal{Q}^a_{\tilde{\mathcal{R}}}(\bm{k}_1)\mathcal{Q}^b_{\tilde{\mathcal{R}}}(\bm{k}_2)\mathcal{Q}^c_{\tilde{\mathcal{R}}}(\bm{k}_3))\rangle
\\\nonumber &\quad+ (2\pi)^3\delta^{(3)}(\bm{k}_1
+\bm{k}_2+\bm{k}_3)\frac{6}{5}f_{NL}\left[\mathcal{N}_a\mathcal{N}_bS^{ab}\right]^2\frac{\tilde{H}^4}{4}\frac{k_1^3
+ k_2^3 + k_3^3}{k_1^3k_2^3k_3^3}.
\end{align} 
Once again, all quantities in the above expressions should be evaluated
at an initial time $\eta_\ast$ that is shortly after the modes in
question have left the horizon.  However, as the $\delta N$ formalism is
by definition independent of $\eta_\ast$, we can evaluate 
them exactly at horizon crossing, i.e. $k = aH$.  
We remark here that under the slow-roll approximation the horizon crossing times in the two frames coincide, i.e.
$k=\mathcal{H}\approx \tilde{\mathcal{H}}$ when $|f'/f| \ll |\cal H|$.

Using \eqref{eq:1}, and following very closely the calculation outlined
in \cite{ns}, to leading order in the slow-roll approximation we
find the spectral index as 
\begin{equation}\label{tiltJ}
n_s-1:= 
\frac{d\ln\mathcal{P}_{\zeta}}{d\ln k}
 =2\frac{d\tilde{H}/d\tilde{t}}{\tilde{H}^2}-\frac{2}{\mathcal{N}_a\mathcal{N}^a}+
 \frac{2\mathcal{N}^a\mathcal{N}^b}{3\tilde{H}^2\mathcal{N}_e\mathcal{N}^e}
 \left[\tilde{\nabla}_a\tilde{\nabla}_b\tilde{V}+\tilde{R}_{acbd}\frac{d\phi^{c}}{d\tilde{t}}\frac{d\phi^{d}}{d\tilde{t}}\right].
\end{equation}
%where $d\tilde{t}=\sqrt{2f}dt=\sqrt{2f}ad\eta$ is the time coordinate in the Einstein frame.

\subsection{Curvature perturbation in the Einstein frame and comparing the frames}

The expansion for $\delta \tilde{N}$ in the Einstein frame follows the
standard procedure, and on making the replacements
$\delta\phi^a_{\tilde{\mathcal{R}}}\rightarrow Q^a_{\tilde{\mathcal{R}}}$ such that
\begin{align}
%\tilde{\zeta} &= \tilde{N}_a\left(Q^a_{\tilde{\mathcal{R}}} - \frac{1}{2}
%{}^{(S)}\Gamma^a_{cd}Q^c_{\tilde{\mathcal{R}}(1)}Q^d_{\tilde{\mathcal{R}}(1)}\right) +
%\frac{1}{2}\tilde{N}_{ab}Q^a_{\tilde{\mathcal{R}}(1)}Q^b_{\tilde{\mathcal{R}}(1)}\\
\tilde{\zeta}&=\tilde{N}_aQ^a_{\tilde{\mathcal{R}}}+
\frac{1}{2}\tilde{\nabla}_a\tilde{\nabla}_b\tilde{N}Q^a_{\tilde{\mathcal{R}}(1)}Q^b_{\tilde{\mathcal{R}}(1)},
\end{align}
we obtain the power spectrum, non-gaussianity and the spectral tilt in
the Einstein frame as
\begin{gather}
%\begin{align}
\label{specE}
\mathcal{P}_{\tilde{\zeta}}(k) =
\tilde{N}_a\tilde{N}_bS^{ab}\left(\frac{\tilde{H}}{2\pi}\right)^2,\qquad
\tilde{f}_{NL} =\frac{5}{6}
\frac{\tilde{N}_a\tilde{N}_b\tilde{\nabla}_c\tilde{\nabla}_d\tilde{N}S^{ac}S^{bd}}{\left[\tilde{N}_e\tilde{N}_{f}S^{ef}\right]^2},
\\
%\mbox{and}\qquad
\label{tiltE}
\tilde{n}_s-1:=
\frac{d\ln\mathcal{P}_{\tilde{\mathcal{\zeta}}}}{d\ln k} 
 = 2\frac{d\tilde{H}/d\tilde{t}}{\tilde{H}^2} - \frac{2}{\tilde{N}^a\tilde{N}_a}
 + \frac{2\tilde{N}^a\tilde{N}^b}{3\tilde{H}^2\tilde{N}_e\tilde{N}^e}
 \left[\tilde{\nabla}_a\tilde{\nabla}_b\tilde{V}+\tilde{R}_{acbd}\frac{d\phi^{c}}{d\tilde{t}}\frac{d\phi^{d}}{d\tilde{t}}\right].
\end{gather}
%\end{align}
The important remaining task, therefore, is to establish the
relationship between derivatives of $N$ and of $\tilde{N}$.  To do this
we look at the integral expressions for $N$ and $\tilde{N}$
\begin{align}\label{N}
&N = \int^{\omega=const.}_{\mathcal{R}=0}\mathcal{H}d\eta,\\\label{Ntilde}
&\tilde{N} =
\int^{\tilde{\omega}=const.}_{\tilde{\mathcal{R}}=0}\tilde{\mathcal{H}}d\eta =
\int^{\tilde{\omega}=const.}_{\tilde{\mathcal{R}}=0}\left(\mathcal{H} +
\frac{f^\prime}{2f}\right)d\eta =
\int^{\tilde{\omega}=const.}_{\tilde{\mathcal{R}}=0}\mathcal{H}d\eta +
\frac{1}{2}\ln{\left(\frac{f_{\tilde{\omega}=const.}}{f_{\tilde{\mathcal{R}}=0}}\right)},
\end{align}
where $\mathcal{R}=0$ and $\tilde{\mathcal{R}}=0$ denote the initial flat slices 
from which we count the number of e-foldings in the Jordan and Einstein frame 
respectively and $\mathcal{\omega}=const.$ and $\tilde{\mathcal{\omega}}=const.$ similarly 
specify the final constant energy surfaces up to which we integrate in the Jordan and Einstein frames respectively.
We see that the integral term on the right-hand side of the Einstein frame expression
\eqref{Ntilde} is the same as \eqref{N} except for the limits of the
integration, and also that there is an additional term in the Einstein
frame expression coming from the variation of the non-minimal coupling
factor $f$.  Let us consider the two limits of the integral terms in
turn.

The lower limit of each integral corresponds to a flat hypersurface at a
time shortly after the scales under consideration have left the horizon, and
we know that in general the flat hypersurfaces as defined in the two
frames do not coincide.  In the expansions for $\delta N$ and $\delta
\tilde{N}$ in terms of the initial field values on the respective flat
hypersurfaces, the difference in definition of the hypersurfaces will
manifest itself in the replacement $\delta\phi^a_{\mathcal{R}}
\rightarrow\delta\phi^a_{\tilde{\mathcal{R}}}$, which we have already taken care
of in the preceding subsection.

The upper limit of each integral corresponds to the surface on which the
final curvature perturbation is evaluated, and any difference between
the upper limits in the two frames will show up as a contribution to the
difference between derivatives of $N$ and $\tilde{N}$ with respect to
the initial conditions.  We know that in the $\delta N$ formalism we
must take the final surface to be one of constant energy in order that
$\delta N$ coincide with $\zeta$.  If we take this constant energy
surface to be at a time when isocurvature modes are still present, then
one finds that constant energy surfaces as defined in the two frames do
not necessarily coincide, thus meaning that $\tilde{N}_a\neq N_a$ etc.
However, if we take the final surface to be one of
constant energy in the radiation dominated phase after inflation, and
assume that no isocurvature modes remain during this epoch, then the  upper
limits of the two integrals coincide, and consequently so too do the
contributions of the integral terms to $N_a$ and $\tilde{N}_a$ etc.

We can show explicitly that in any adiabatic limit the two expansions for
$\delta N$ and $\delta \tilde{N}$ do coincide, as we would expect.  We
have already argued that as the final surface is in an epoch where
the two frames coincide, its effect on the values of $N_a$ and
$\tilde{N}_a$ etc is frame independent.  One might still be concerned
that the difference in the initial flat hypersurfaces as defined in the
two frames might lead to a difference in the expansions for $\delta N$
and $\delta \tilde{N}$.  What we find, however, is that this difference
in definition of the initial hypersurface is exactly cancelled by the
additional log term in the Einstein frame expression for $\tilde{N}$.
Let us see this explicitly up to second order.

Using \eqref{N} and \eqref{Ntilde} we see that derivatives of
$\tilde{N}$ are related to those of $N$ as
\begin{align}
\tilde{N}_a &= N^{\tilde{\omega}}_a - \left.\frac{f_a}{2f}\right|_\ast +
\left.\frac{f_b}{2f}\right|_\diamond\frac{\partial\phi^b_{\tilde{\omega}}}{\partial\phi^a_\ast},\\
\tilde{N}_{ab} &= N^{\tilde{\omega}}_{ab} -
\left.\frac{f_{ab}}{2f}\right|_\ast+\left.\frac{f_af_b}{2f^2}\right|_\ast
+
\left.\frac{f_{cd}}{2f}\right|_\diamond\frac{\partial\phi^c_{\tilde{\omega}}}{\partial\phi^a_\ast}\frac{\partial\phi^d_{\tilde{\omega}}}{\partial\phi^b_\ast}-\left.\frac{f_cf_d}{2f^2}\right|_\diamond\frac{\partial\phi^c_{\tilde{\omega}}}{\partial\phi^a_\ast}\frac{\partial\phi^d_{\tilde{\omega}}}{\partial\phi^b_\ast}+\left.\frac{f_c}{2f}\right|_\diamond\frac{\partial^2\phi^c_{\tilde{\omega}}}{\partial\phi^a_\ast\partial\phi_\ast^b},
\end{align}
where the superscript ${\tilde{\omega}}$ on derivatives of $N$ indicates that
they are derivatives of the Jordan frame e-folding number but with the
upper limit of the integral being that defined in the Einstein frame.
As such, combining with \eqref{Jzeta}, we find that the difference
between $\zeta$ and $\tilde{\zeta}$ is
\begin{align}\label{eq:15}
\zeta - \tilde{\zeta} &= (N_a - N^{\tilde{\omega}}_a)\delta\phi^a_{\tilde{\mathcal{R}}}
+ \frac{1}{2}(N_{ab} -
N^{\tilde{\omega}}_{ab})\delta\phi^a_{\tilde{\mathcal{R}}(1)}\delta\phi^b_{\tilde{\mathcal{R}}(1)}
\\\nonumber
&\quad-\left.\frac{f_b}{2f}\right|_\diamond\frac{\partial\phi^b_{\tilde{\omega}}}{\partial\phi^a_\ast}\delta\phi^a_{\tilde{\mathcal{R}}}
-\left(\left.\frac{f_{cd}}{2f}\right|_\diamond\frac{\partial\phi^c_{\tilde{\omega}}}{\partial\phi^a_\ast}\frac{\partial\phi^d_{\tilde{\omega}}}{\partial\phi^b_\ast}-\left.\frac{f_cf_d}{2f^2}\right|_\diamond\frac{\partial\phi^c_{\tilde{\omega}}}{\partial\phi^a_\ast}\frac{\partial\phi^d_{\tilde{\omega}}}{\partial\phi^b_\ast}+\left.\frac{f_c}{2f}\right|_\diamond\frac{\partial^2\phi^c_{\tilde{\omega}}}{\partial\phi^a_\ast\partial\phi_\ast^b}\right)\delta\phi^a_{\tilde{\mathcal{R}}(1)}\delta\phi^b_{\tilde{\mathcal{R}}(1)}.
\end{align}   
If we assume that an adiabatic limit is reached then we have two
simplifications.  Firstly, as discussed above, the definition of the
final surface becomes frame independent, meaning $N_a =
N^{\tilde{\omega}}_a$ and $N_{ab} = N^{\tilde{\omega}}_{ab}$.  Secondly, the final
values of all the
fields are independent of the initial conditions,
i.e. $\partial\phi^a_{\tilde{\omega}}/\partial\phi^b_\ast = 0$.  From the above
expression it is clear that in this case $\zeta =\tilde{\zeta}$.  With
regard to the expressions for the power spectra, spectral tilts and $f_{NL}$
parameters in the two frames, their equivalence in an adiabatic limit is
evident from the fact that in such a limit $\tilde{N}_a=\mathcal{N}_a$
and $\tilde{\nabla}_a\tilde{\nabla}_b\tilde{N}=\tilde{\nabla}_a\tilde{\nabla}_b\mathcal{N}$.  

More
generally, we see that even in the case that an adiabatic limit is not
reached, the difference in definition of the initial flat hypersurfaces
does not affect $\zeta - \tilde{\zeta}$.  This is because the
differences resulting from $\delta\phi^a_{\mathcal{R}} \leftrightarrow
\delta\phi^a_{\tilde{\mathcal{R}}}$ exactly cancel with the additional
$\ln(f_{\tilde{\mathcal{R}}=0})/2$ term coming from \eqref{Ntilde}.
    
\section{Analytically soluble models}\label{analsec}

Having found general expressions for the non-linear curvature perturbation in both frames, 
in this section we consider cases where the number of e-foldings and its dependence on 
the initial field values can be determined analytically.  

%and apply the results to analytically soluble models and 
%discuss in detail the properties of spectral quantities and their frame dependence.
%In order to get some idea as to how the curvature perturbations $\zeta$
%and $\tilde{\zeta}$ are related in the case that isocurvature modes are
%still present in the final epoch, in this section we consider
%some analytically soluble examples.

\subsection{Solubility conditions}

Varying the Jordan frame action \eqref{JAc}
%in the Jordan frame
with respect to the fields $\phi^a$ and metric $g_{\mu\nu}$, at background level %of the metric \eqref{metDecomp}
the equations of motion and
Friedmann equation are given as
\begin{align}
&\frac{D\dot{\phi}^a}{dt} +3H\dot{\phi}^a + h^{ab}\left(V_{b} -
f_{b}R\right) = 0\quad\quad\mbox{and}\quad\quad3H^2 = \frac{1}{2
f}\Big(\frac{1}{2}h_{ab}\dot{\phi}^a\dot{\phi}^b+V-6H\dot{f}\Big),
\end{align}
where $D/dt$ is the covariant derivative with respect to $h_{ab}$ defined such that
$D\dot{\phi}^a/dt = \ddot{\phi}^a+{}^{(h)}\Gamma^{a}_{bc}\dot{\phi}^b\dot{\phi}^c$.  We would like to approximate these equations as
\begin{align}\label{jFsR}
&3H\dot{\phi}^a \simeq - h^{ab}\left(V_{b} - 2V \frac{f_{b}}{f}\right)
= -f^2 h^{ab}W_{b}\quad\quad\mbox{and}\quad\quad3H^2 \simeq \frac{V}{2
f}
\end{align}
in the slow-roll limit, where $W = V/f^2$
(see Eq. (\ref{albert})),
and a dot denotes taking the derivative with respect to the 
physical time in the Jordan frame.\footnote{Note that here we use the Jordan frame equations
of motion to determine the derivatives of $N$ with respect to the
initial conditions and we then relate these to derivatives of
$\tilde{N}$.  We could equally use the Einstein frame equations of
motion to determine derivatives of $\tilde{N}$ and then relate these
back to derivatives of $N$.
 We discuss this further in appendix \ref{EFappend}.} 
 For this approximation to be
valid we require
\begin{align}\label{jFcR}
\frac{D\dot{\phi}^a}{dt}\ll H\dot{\phi}^a,\quad\quad
h_{ab}\dot{\phi}^a\dot{\phi}^b \ll V \quad\quad\mbox{and}\quad\quad
H\dot{f} \ll V.
\end{align}
Using the first of \eqref{jFsR} to determine $\frac{D\dot{\phi}^a}{dt}$
we find
\begin{equation}
\frac{1}{H\dot{\phi}^a}\frac{D\dot{\phi}^a}{dt} \simeq
-\frac{\dot{H}}{H^2} -
\frac{\dot{\phi}^b\nabla_b\left(f^2 h^{ac}W_{c}\right)}{3H^2\dot{\phi}^a}\simeq
-\frac{\dot{H}}{H^2} -
\frac{2f h^{bd}W_{d}\nabla_b\left(f^2 h^{ac}W_{c}\right)}{V h^{ae}W_{e}},
\end{equation}
%where we have defined $W_{eff}^{,a} = f^2 h^{ab}W_{b}$.  
where $\nabla_a$ is the usual covariant derivative with respect to $h_{ab}$.  Note that here
the index $a$ is not summed over.  Similarly we find
\begin{align}
\frac{h_{ab}\dot{\phi}^a\dot{\phi}^b}{V} \simeq
\frac{2f^5 h^{ab}W_{a}W_{b}}{3V^2}\quad\quad\mbox{and}\quad\quad\frac{H\dot{f}}{V}\simeq
\frac{\dot{f}}{6Hf}\simeq-\frac{f^2h^{ab}f_{a} W_{b}}{3V},
\end{align}   
so that we are able to define the slow-roll conditions
%\begin{align}\label{sR1}
%&\epsilon = \frac{fh_{ab}W_{eff}^{,a}W_{eff}^{,b}}{V^2} \ll
%1\\\label{sR2} &\eta^{(a)} =
%\frac{2fW_{eff}^{,b}\nabla_bW_{eff}^{,a}}{VW_{eff}^{,a}} \ll
%1\\\label{sR3} &\delta = \frac{f_{a}W_{eff}^{,a}}{3V} \ll 1.
%\end{align}
\begin{align}\label{sR1}
\epsilon &\equiv \frac{fh^{ab}W_{a}W_{b}}{W^2}\quad ; \quad\epsilon \ll 1\\\label{sR2}
\eta^{(a)} &\equiv 2f\frac{h^{bc}W_{c}\nabla_b\left(h^{ad}W_{d}\right)}{Wh^{ae}W_{e}}+\frac{4h^{bc}f_{b}W_{c}}{W}\quad ; \quad |\eta^{(a)}| \ll 1\\\label{sR3}
\delta &\equiv \frac{h^{ab}f_{a}W_{b}}{3W}\quad ; \quad |\delta|\ll 1.
\end{align}
Note that to derive these conditions we have also assumed $\dot{H}\ll
H^2$.  Taking the time derivative of the second of \eqref{jFsR} we find
\begin{equation}
\frac{\dot{H}}{H^2} = -\epsilon - 3\delta,
\end{equation} 
so that if the slow roll conditions \eqref{sR1} and \eqref{sR3} are
satisfied then $\dot{H}\ll H^2$ will also be satisfied.

Having established the slow-roll equations and corresponding consistency
relations we now look to determine under what circumstances we are able
to solve for the number of e-folds of inflation analytically.  We start
by replacing cosmic time $t$ with the number of e-folds using $dN = Hdt$.  The first of \eqref{jFsR} can
then be written as
\begin{equation}
\label{eom_sR}
\frac{d\phi^a}{dN} = -2fh^{ab}\frac{W_{b}}{W}.
\end{equation}
This can be solved analytically if
\begin{equation}
\label{analCon}
2fh^{ab}\frac{W_{b}}{W} = \frac{g^{(a)}(\phi^a)}{F({\bm \phi})},
\end{equation} 
where $g^{(a)}(\phi^a)$ represents some function of just the $a$'th field
and $F({\bm \phi})$ is some function of all the fields, as this allows
us to write
\begin{equation}
\frac{1}{g^{(a)}(\phi^a)}\frac{d\phi^a}{dN} = -\frac{1}{F({\bm \phi})}
\end{equation}
for all $a$, which in turn means that any one field can be expressed as
a function of any one of the other fields.  Note that here the index $a$
is not summed over.  Explicitly, we can see that \eqref{analCon} is
satisfied if
\begin{equation}\label{solh}
h^{ab} = \frac{1}{G({\bm
\phi})}\rm{diag}\left(h^{(1)}(\phi^1),\,h^{(2)}(\phi^2),\,...,\,h^{(n)}(\phi^n)\right)
\end{equation}
and either
\begin{equation}
W = \prod_aW^{(a)}(\phi^a)\quad\quad\mbox{or}\quad\quad W = \sum_a
W^{(a)}(\phi^a),
\end{equation}
where again $h^{(a)}(\phi^a)$ and $W^{(a)}(\phi^a)$ represent functions of the
single field $\phi^a$ and $G({\bm \phi})$ is some function of all the
fields.  In fact, the form of the field-space metric \eqref{solh} can
be simplified further by noticing that the $h^{(a)}(\phi^a)$ can always be absorbed
by a field redefinition.  As such, making the additional simplifying
assumption $G(\bm{\phi}) = 1$, we see that $F(\bm{\phi}) = 1/(2f)$ and
$g^{(a)}(\phi^a) = W_{a}/W$.

Following \cite{multibrid},\footnote{See also \cite{separable1,separable2,separable3,separable4} for considerations of
analytically soluble applications of the $\delta N$ formalism in
multi-field inflation.} we next introduce the
coordinates $q^a=qn^a$ defined by 
\begin{equation}
 \ln q^a = \int\frac{d\phi^a}{g^{(a)}(\phi^a)}\qquad\mbox{and}\qquad\sum_a(n^a)^2=1.
\end{equation}
The equations of motion can then be expressed as
\begin{equation}
 \frac{d\ln q}{dN}=-\frac{1}{F(q,n^a)}\qquad\mbox{and}\qquad\frac{dn^a}{dN}=0,
\end{equation} 
and on integrating the first of these we obtain
\begin{equation}
 N = -\int^\diamond_\ast Fd\ln q = \int^\ast_\diamond Fd\ln q.
\end{equation}
We can now proceed to calculate the derivatives of $N$ with
respect to the initial field values.  

\subsection{The product separable case}

In this paper we focus on the product separable case where $W = \prod_aW^{(a)}(\phi^a)$,
for which we find \cite{separable2}
\begin{equation}
N_c =F_\diamond\frac{g_\diamond^{(c)}}{g_\ast^{(c)}}\frac{\partial \omega/\partial
\phi_\diamond^c}{\sum_a\frac{\partial \omega}{\partial \phi_\diamond^a} g_\diamond^{(a)}} +
\int^{\ast}_{\diamond}\frac{\partial F}{\partial
\phi^c}\frac{g^{(c)}}{g^{(c)}_\ast} d\ln q,
\end{equation}
where $\omega(\bm{\phi}_\diamond) = \rm{const.}$ is the condition specifying the
final surface up to which the number of e-folds is calculated.

If we take the final surface in
each frame to correspond to a constant energy surface, i.e. $\omega=\rho =
3H^2 = fW/2={\rm const.}$ in the Jordan frame and $\tilde{\omega}=\tilde{\rho} =
3\tilde{H}^2 = W/4= {\rm const.}$ in the Einstein frame, then we have
\begin{equation}
N_c
=\frac{1}{2}\frac{g_\diamond^{(c)}}{g_\ast^{(c)}}\frac{\left(\left.\frac{f_{c}}{f}\right|_\diamond +
g_\diamond^{(c)}\right)}{\epsilon_\diamond +3\delta_\diamond} +
\int^{\ast}_{\diamond}\frac{\partial F}{\partial
\phi^c}\frac{g^{(c)}}{g^{(c)}_\ast} d\ln q
\end{equation}
in the Jordan frame, and the contribution $N^{\tilde{\omega}}_c$ to
$\tilde{N}_c$ in the Einstein frame is
\begin{align}
N^{\tilde{\omega}}_c
&=\frac{1}{2}\frac{(g_\diamond^{(c)})^2}{\epsilon_\diamond g_\ast^{(c)}} + \int^{\ast}_{\diamond}\frac{\partial F}{\partial
\phi^c}\frac{g^{(c)}}{g^{(c)}_\ast} d\ln q.
\end{align}

The
above expression can of course be differentiated again to find
higher-order derivatives of $N$.  The second order derivatives can be
found as 
\begin{align}\label{eq:23}
 N_{cd} &= \sum_i\left(\frac{g_\diamond^{(d)}}{g_\ast^d}\delta^i_d 
   - g_\diamond^i\frac{g_\diamond^{(d)}}{g_\ast^{(d)}}\frac{\partial \omega/\partial\phi_\diamond^d}{\sum_a g_\diamond^{(a)} \partial \omega/\partial\phi^a_\diamond}\right)\left[\left(\frac{\partial F_\diamond}{\partial\phi^i_\diamond}\frac{g^{(c)}_\diamond}{g^{(c)}_\ast}
          + \frac{F_\diamond}{g^{(c)}_\ast}\frac{\partial g^{(c)}_\diamond}{\partial\phi^i_\diamond}\right)\frac{\partial \omega/\partial\phi^c_\diamond}{\sum_a g_\diamond^{(a)}\partial \omega/\partial\phi_\diamond^a}\right.\\\nonumber
          &\quad\left.\qquad+ F_\diamond\frac{g^{(c)}_\diamond}{g^{(c)}_\ast}\frac{1}{\sum_a g_\diamond^{(a)}\partial \omega/\partial\phi_\diamond^a}
            \left\{\frac{\partial^2
 \omega}{\partial\phi^i_\diamond\partial\phi^c_\diamond}-\frac{\partial \omega/\partial\phi^c_\diamond}{\sum_a g_\diamond^{(a)}\partial \omega/\partial\phi_\diamond^a}\sum_a\left(g^{(a)}_\diamond\frac{\partial^2\omega}{\partial\phi^i_\diamond\partial\phi_\diamond^a}+\frac{\partial g^{(a)}_\diamond}{\partial \phi^i_\diamond}\frac{\partial \omega}{\partial\phi^a_\diamond}\right)\right\}\right]\\\nonumber
&\quad-F_\diamond\frac{g^{(c)}_\diamond}{(g^{(c)}_\ast)^2}\frac{\partial g_\ast^{(c)}}{\partial\phi^d_\ast}\frac{\partial \omega/\partial\phi^c_\diamond}{\sum_a g_\diamond^{(a)}\partial \omega/\partial\phi_\diamond^a}+\left[\frac{\partial F}{\partial\phi^c}\frac{g^{(c)}}{g^{(c)}_\ast}\right]_\diamond\frac{g^{(d)}_\diamond}{g_\ast^{(d)}}\frac{\partial \omega/\partial\phi^d_\diamond}{\sum_a g_\diamond^{(a)}\partial \omega/\partial\phi_\diamond^a}\\\nonumber
&\quad+\frac{1}{g_\ast^{(c)}}\int^\ast_\diamond\frac{\partial}{\partial\phi^d}\left(\frac{\partial F}{\partial\phi^c}g^{(c)}\right)\frac{g^{(d)}}{g_\ast^{(d)}}d\ln q^a
-\frac{\partial g_{\ast}^{(c)}/\partial\phi^d_\ast}{(g_\ast^{(c)})^2}\int^\ast_\diamond\frac{\partial F}{\partial\phi^c}g^{(c)}d\ln q^a.
\end{align}
Note that in the above expression, as $g^{(a)}_\diamond$ is only a function of $\phi^a_\diamond$ and 
$g^{(a)}_\ast$ only a function of $\phi^a_\ast$, the terms involving $\partial g^{(c)}_\ast/\partial\phi^d_{\ast}$ are 
only non-zero for $c=d$ and similarly those of the form $\partial g^{(a)}_\diamond/\partial\phi^i_\diamond$
are only non-zero for $i=a$.  

As a final comment before considering some examples we note that the
validity of \eqref{eq:23} is limited to cases where the curvature of the
Einstein frame field-space metric can be neglected when differentiating
the equations of motion.  For details regarding this issue see
appendix \ref{slowrollapp}.  Here we simply present the most relevant subset of
the sufficient conditions for slow-roll 
and neglection of the
field-space curvature to be valid approximations:

\begin{equation}
\sqrt{f}\frac{W_a}{W}\sim\mathcal{O}(\epsilon^{1/2}),\quad
\frac{fW_{ab}}{W}\sim\mathcal{O}(\epsilon),\quad 
\frac{f_a}{\sqrt{f}}\sim\mathcal{O}(\epsilon^{1/2}),\quad
f_{ab}\lesssim\mathcal{O}(\epsilon)\quad\mbox{and}\quad
  \sqrt{f}f_{abc}\lesssim\mathcal{O}(\epsilon^{1/2}).  
\end{equation}
  
\section{Non-minimally coupled spectator field}\label{nmcssec}

As a first example, in this section we consider a 
model involving a non-minimally coupled spectator field.  Here, by `spectator' 
field we mean a field which is non-dynamical at background level.
 However, as we will see, it does contribute to the curvature perturbation throughout 
 inflation, meaning that it is not a spectator field in the usual sense.

\subsection{The model and derivatives of $\tilde{N}$ and $N$}

The model is similar to that considered in
\cite{wms}, where we have a minimally coupled field $\phi$ that drives
inflation and a spectator field $\chi$ that is non-minimally coupled,
i.e. $f = f(\chi)$.  As mentioned above, by `spectator' field we mean that at background level 
$\chi^\prime=0\rightarrow W_\chi = g^{(\chi)}=0$.  Note that because the
$\chi$ field is non-dynamical, the constraints on the time variation of
$f$, e.g. $\delta\ll 1$, do not put any constraints on $f_\chi$ or
$f_{\chi\chi}$.  We will thus have to rely on observables to constrain
these quantities instead.  Full details regarding the derivatives of
$\tilde{N}$ and $N$ with respect to $\phi$ and $\chi$ can be found in
appendix \ref{nmcsapp}, and here we simply discuss the main results.

The key feature of this model is that, despite being non-dynamical at
background level, fluctuations of the $\chi$ field contribute to
the curvature perturbation as a result of its non-minimal coupling.  The
details of this contribution are encoded in the derivatives of
$\tilde{N}$ and $N$ with respect to $\chi$, which are given as
\begin{align}
 \tilde{N}_\chi &= -\frac{1}{2}\int^\ast_\diamond \frac{f_{\chi}}{f^2}
\frac{1}{g^{(\phi)}}d\phi,\\
\tilde{N}_{\chi\chi}
&=\frac{1}{2\epsilon_\diamond}\frac{W_{\chi\chi}}{W}-\frac{1}{2}\int^\ast_\diamond\left(\frac{f_{\chi\chi}}{f^2}-\frac{2(f_{\chi})^2}{f^3}\right)\frac{1}{g^{(\phi)}}d\phi,\\
N_\chi &=\tilde{N}_\chi +
 \frac{f_{\chi}}{2f_\diamond\epsilon_\diamond},\\
N_{\chi\chi} &= \tilde{N}_{\chi\chi} +\frac{1}{2\epsilon_\diamond}\left(\frac{f_{\chi\chi}}{f} 
-2\frac{(f_\chi)^2}{f\epsilon_\diamond}\frac{W_{\chi\chi}}{W}-
\frac{(f_{\chi})^2}{f^2}\left(5-\frac{\eta^{(\phi)}_\diamond}{\epsilon_\diamond}\right)\right).
\end{align}
Looking at these expressions, the first thing to notice is that
$\tilde{N}_\chi\ne N_\chi$ and $\tilde{N}_{\chi\chi}\ne N_{\chi\chi}$,
which will subsequently mean that $\zeta\ne\tilde{\zeta}$.
Furthermore, seeing as the differences involve
factors of $\epsilon_\diamond^{-1}$, we might expect them to be
significant at early times when the slow-roll condition is satisfied.
As we approach the end of inflation, however, where the slow-roll approximation breaks down
and $\epsilon_\diamond$ correspondingly approaches one, we might expect
the differences to be less significant.\footnote{We must, of course, be wary of trusting our results at these later times, as our analytic results rely on the validity of the slow-roll approximation.}  At linear order we find 
\begin{equation}\label{eq:6}
\zeta - \tilde{\zeta} = \left(\frac{f_\chi}{2f\epsilon_\diamond} -
\frac{f_\chi}{2f}\right)\delta\chi_{\tilde{\mathcal{R}}}\simeq
\frac{f_\chi}{2f\epsilon_\diamond} \delta\chi_{\tilde{\mathcal{R}}},
\end{equation}
the form of which is in agreement with the result found using linear perturbation
theory in \cite{wms}.  From this expression we see that in order for the
two curvature perturbations to be of the same order of magnitude we
require $f_\chi/\sqrt{f}\lesssim\mathcal{O}(\epsilon^{1/2})$.  Another
point we notice is the integral nature of $\tilde{N}_\chi$ and
$\tilde{N}_{\chi\chi}$.  This suggests that the contribution of the
$\chi$ field to the curvature perturbation will be an increasing
function of time, and we might therefore expect it to affect the tilt of
the power spectrum.  In this section we will assume that soon after
the instance $\epsilon_{\diamond}=1$ the fields quickly decay into
radiation and the adiabatic limit is reached.  Then, the spectral
properties at $\epsilon_{\diamond}=1$ can be naturally interpreted as
those we observe.

% \subsection{Concrete example with $V^{(\phi)}(\phi)=m^2\phi^{2p}$}
 
In order to allow us to analyse more concretely the contribution of the
non-minimally coupled $\chi$ field to the curvature perturbation, let us
now focus on a particular form of potential for the inflaton field
$\phi$, taking $V=V^{(\chi)}(\chi)V^{(\phi)}(\phi)$ and
$V^{(\phi)}(\phi)=m^2\phi^{2p}$.  The corresponding expressions for the
derivatives of $\tilde{N}$ and $N$ can be found in appendix \ref{nmcsapp}.  

%\subsubsection
\subsection{Power spectrum and tilt}
\begin{figure}
\centering
\begin{minipage}{.5\textwidth}
  \centering
  \includegraphics[width=\linewidth]{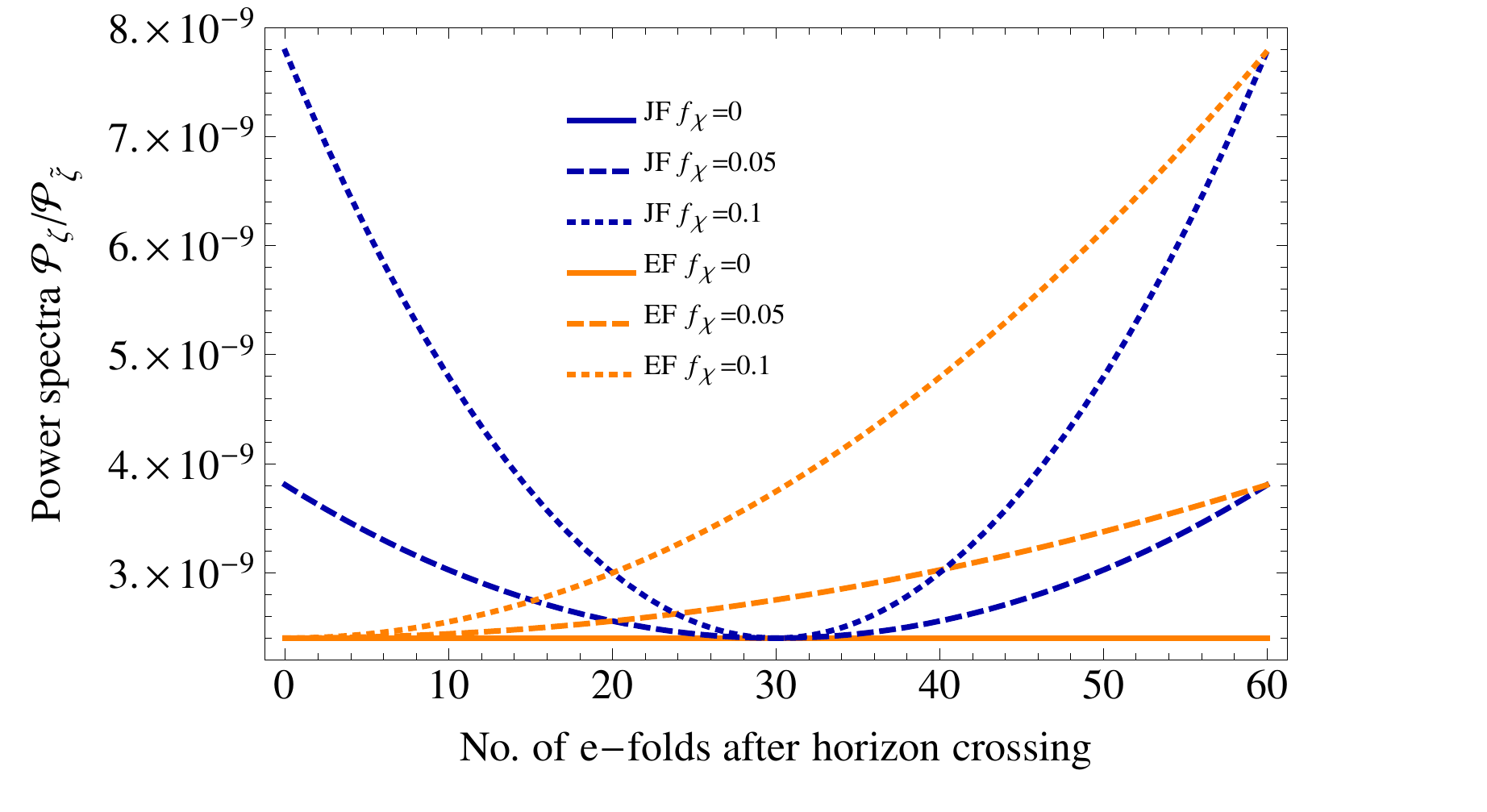}
  %\caption*{(a)}
  %\label{fig:test1}
\end{minipage}%
\begin{minipage}{.5\textwidth}
  \centering
  \includegraphics[width=\linewidth]{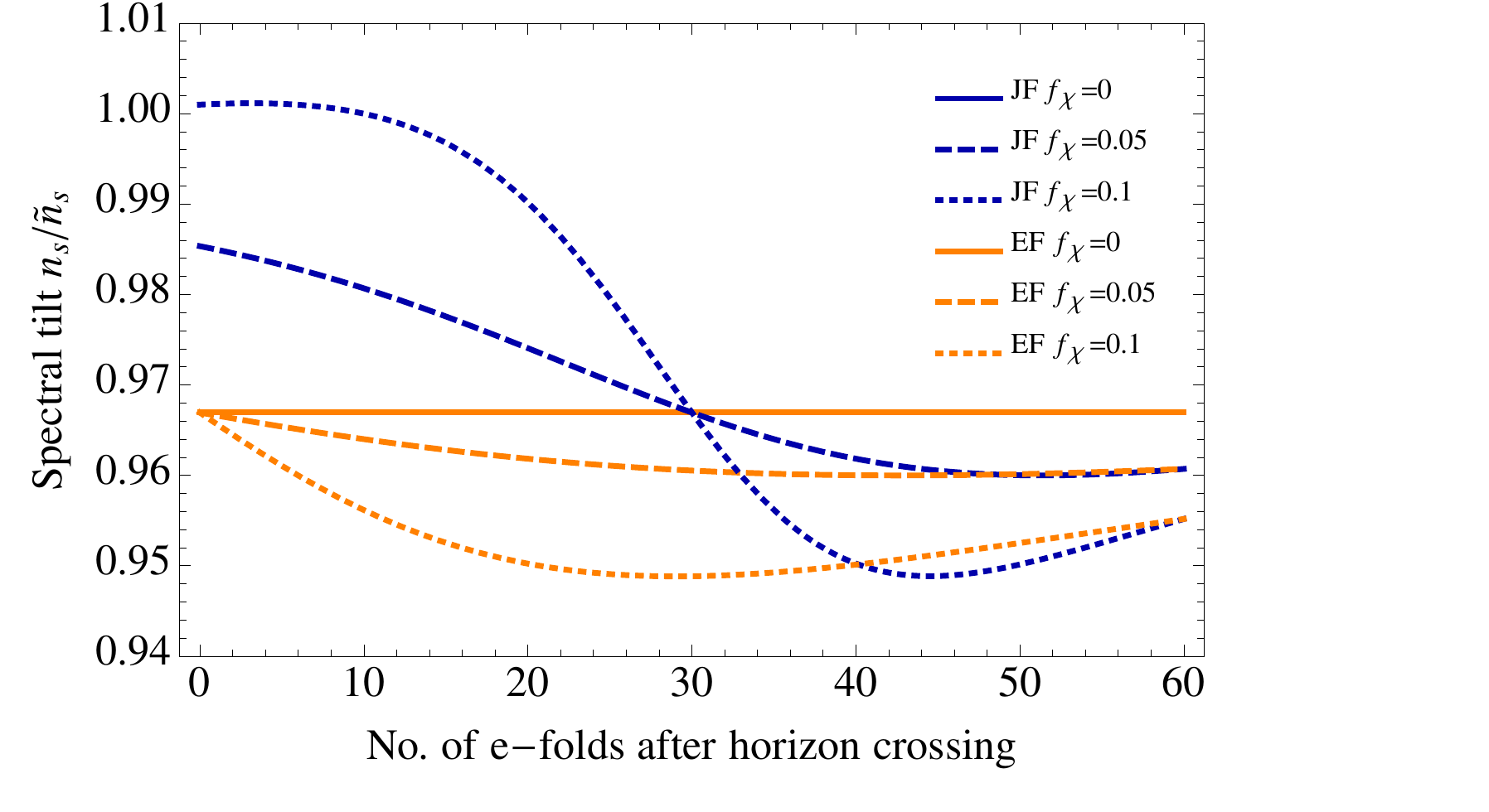}
  %\caption*{(b)}
  %\label{fig:test2}
\end{minipage}%
%\begin{minipage}{.33\textwidth}
%  \centering
%  \includegraphics[width=.9\linewidth]{fNL.pdf}
%  \caption*{(c)}
%  %\label{fig:test2}
%\end{minipage}
\caption{Evolution of the power spectrum (left) and spectral tilt (right) in the
 Jordan (JF) and Einstein (EF) frames for a range of $f_\chi$.  Here we have set
 $2f=1$, $V^{(\chi)}=1$, $f_{\chi\chi}=W_{\chi\chi}=0$, $N_\ast=60$, $p=1$
 and $m^2=1.94\times10^{-11}$.}
\label{fig:1}
\end{figure}

In
figure \ref{fig:1} we have plotted the Jordan and Einstein frame power
spectra and tilts as functions of $N$ for a range of values of $f_\chi$.
The end of inflation is taken to be defined by $\epsilon_\diamond = 1$,
and we consider modes that leave the horizon 60 e-folds before the end
of inflation.  For definiteness, the remaining parameters are taken as
follows: $2f=1$ (so that the effective Planck mass is in agreement with
the current value), $f_{\chi\chi}=W_{\chi\chi}=0$,
$V^{(\chi)}(\chi)=1$ and $p=1$.  Lastly, we take $m^2=1.94\times
10^{-11}$ in order that for $f_\chi=0$ the power spectrum is normalised
to the observed $2.4\times 10^{-9}$.

The behaviour of the power spectra and tilts in figure \ref{fig:1} can
be well understood in line with the general arguments given
above.  In the Einstein frame, from \eqref{specE},
we find the power spectrum to be given as 
\begin{equation}
 \mathcal{P}_{\tilde{\zeta}}=\left[\frac{N_\ast}{2pf}S^{\phi\phi}+\left(\frac{f_\chi}{f}\right)^2N^2S^{\chi\chi}\right]\left(\frac{\tilde{H}}{2\pi}\right)^2,
\end{equation}
where $N_\ast$ corresponds to the number of e-foldings before the end of
inflation that the modes under consideration leave the horizon.  In the above equation and in all of the following analysis we will assume $N_\ast\gg p$.  The
first term in the square brackets corresponds to the time independent
contribution from the $\phi$ field, whilst the second term to the time
dependent contribution from the $\chi$ field.  Thus, at $N=0$ the
Einstein frame power spectrum agrees with that in the case that
$f_\chi=0$, but it increases as $N^2$ for $N>0$.  In the Jordan frame, from \eqref{eq:1}, we find
\begin{equation}\label{eq:18}
 \mathcal{P}_{\zeta}=\left[\frac{N_\ast}{2pf}S^{\phi\phi}+\left(\frac{f_\chi}{pf}\right)^2\left(N_\ast-N(1+p)\right)^2S^{\chi\chi}\right]\left(\frac{\tilde{H}}{2\pi}\right)^2.
\end{equation}
Once again, the first term in the square brackets corresponds to the
contribution from the $\phi$ field, whilst the second to that from the
$\chi$ field.  At $N=0$ we see that the contribution from the $\chi$
field has an additional term of the form $(f_\chi N_\ast/fp)^2$ in
comparison with the Einstein frame case.  At the time
$N=N_\ast/(1+p)$ the contribution from the $\chi$ field vanishes,
meaning that the power spectra coincides with that in the case $f_\chi =0$.  In
the case that $p=1$, which is the case in our plot, this time
corresponds to $N=30$, and we also see that the behaviour of
$\mathcal{P}_\zeta$ is symmetric about $N=30$.  
At $N=60$ the power spectra in the two frames coincide to leading order.

Turning to the spectral index, in the Einstein frame, from \eqref{tiltE}, we find
\begin{equation}
 \tilde{n}_s-1 = -\frac{p}{N_\ast}-A\frac{1}{N_\ast}-A\frac{2p}{N_\ast}\frac{S^{\chi\chi}}{f}\frac{f_\chi^2}{f}N,
\end{equation} 
where we have taken $f_{\chi\chi}=0$ (which allows us to neglect the
contribution from the Riemann curvature term),
$W_{\chi\chi}=0$, and
\begin{equation}
 A=\frac{\frac{N_\ast}{2fp}}{\frac{N_\ast}{2fp}+\frac{S^{\chi\chi}}{2f^2}\frac{f_\chi^2}{f}N^2}.
\end{equation}  
For $N=0$ the spectral tilt coincides with that in the case that $f_\chi=0$,
i.e. the third term vanishes and $A=1$.  As $N$ increases, however, we have
two effects: there is an additional redshift coming from the
third term and the contribution to the redshift from both the second and third terms is
suppressed by the factor $A<1$.  For our choice of parameters we can see
from figure \ref{fig:1} that the combination of these effects is to make
the spectrum more red tilted than in the case that $f_\chi=0$, but that
the suppression by $A$ at large $N$ causes the additional red tilt to decrease.  In
the Jordan frame, from \eqref{tiltJ}, we can similarly write      
\begin{equation}
 n_s-1 = -\frac{p}{N_\ast}-B\frac{1}{N_\ast}+B\frac{2S^{\chi\chi}f_\chi^2}{N_\ast f^2}\left(N_\ast-N(1+p)\right),
\end{equation} 
where again we have taken $f_{\chi\chi}=W_{\chi\chi}=0$ and
\begin{equation}
 B=\frac{\frac{N_\ast}{2fp}}{\frac{N_\ast}{2fp}+\frac{S^{\chi\chi}}{2p^2f^2}\frac{f_\chi^2}{f}\left(N_\ast-N(1+p)\right)^2}.
\end{equation}
At $N=0$, as $B<1$ and the third term gives an additional positive
contribution, we see that the spectral tilt will be bluer than that in the case
where $f_\chi=0$.  Then, at $N=N_\ast/(1+p)$, we see that the effect of
the $\chi$ field vanishes and we recover the canonical single field
result.  Finally, at $N=N_\ast$, the result coincides with that in the
Einstein frame.  All these features are clear in figure \ref{fig:1}.

\begin{figure}
\centering\includegraphics[width=\textwidth]{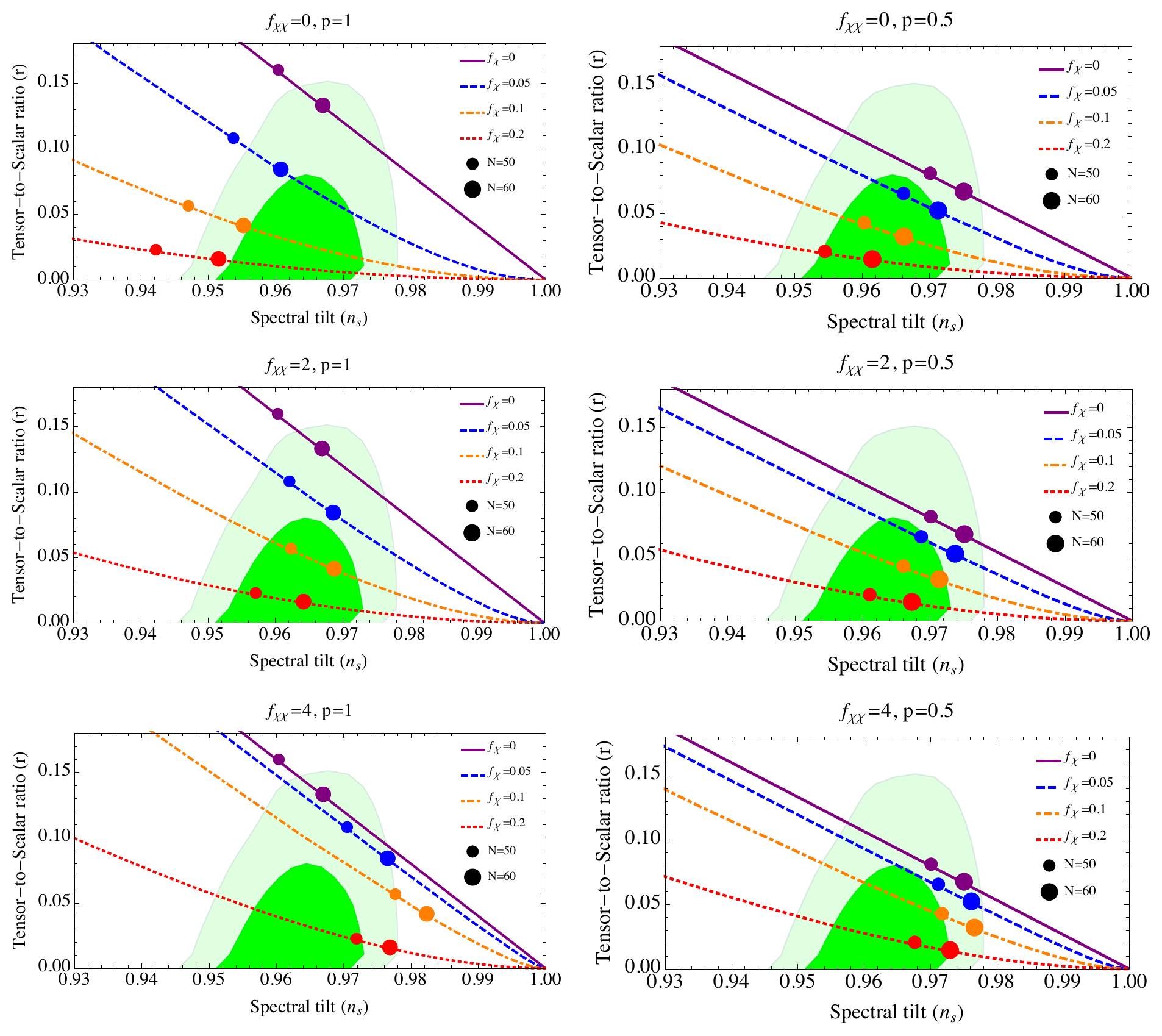}]
\caption{Predictions in the $n_s$-$r$ plane for models with inflaton
 potential of the form $V^{(\phi)}=m^2\phi^{2p}$.  Parameters are as
 specified on each plot.  Additionally, $V^{(\chi)}=2f=1$,
 $W_{\chi\chi}=0$ and $m$ is normalised for each curve such that
 $\mathcal{P}_\zeta = 2.4\times10^{-9}$.  As the Jordan and Einstein
 frames agree at leading order at the end of inflation, here we only plot
 the Jordan frame quantities.  The 68\% and 95\% confidence contours
 from the recent {\it Planck} results have also been plotted.}
\label{fig:2} 
\end{figure}

Having discussed the evolution of the power spectra and spectral tilts, we now
focus on their final values and the dependence of these values on the
non-minimal coupling function.  Allowing for a non-zero $f_{\chi\chi}$,
but still taking $W_{\chi\chi}=0$, in figure \ref{fig:2} we have plotted
predictions in the $n_s$-$r$ plane for a range of $p$, $f_\chi$,
$f_{\chi\chi}$ and $N_\ast$.  As we expect the final Jordan and Einstein frame results to
agree at leading order, we choose to only plot the Jordan frame
parameters.  In each case we choose $m$ such that the
$\mathcal{P}_\zeta$ is normalised to $2.4\times 10^{-9}$, and the
remaining parameters are set as $2f=V^{(\chi)}=1$.

The first feature we notice, which is common to all the plots, is that
as $f_\chi$ is increased the tensor-to-scalar ratio is suppressed.  
This can be easily understood if we recall that tensor and scalar modes are decoupled at linear order
and also that the tensor amplitude is frame-independent, allowing us to simply evaluate it in the Einstein frame.  As a result, the tensor power spectrum takes the standard form $\mathcal{P}_t=8(\tilde{H}/(2\pi))^2$, so the only effect of the additional non-minimally coupled field is any contribution it makes to $\tilde{H}$.  The scalar power spectrum, on the other hand, acquires an additional contribution corresponding to the second term in the square brackets in \eqref{eq:18}.  As such, the tensor-to-scalar ratio is suppressed.  In fact, note that in our specific model, where $2f =V^{(\chi)}=1$ and $\chi^\prime=0$ at background level, $\tilde{H}$ is not modified by the presence of the $\chi$ field, meaning that the tensor power spectrum itself is the same as that in the standard single-field case.  Additionally, as can
be seen in the two plots with $f_{\chi\chi}=0$, a non-zero $f_\chi$
causes the spectrum to become more red-tilted.  In the case $p=0.5$,\footnote{See  \cite{axionmono} for the original single field model with $p=0.5$.}
the combination of these two effects brings the predictions of this
model within the 68\% confidence contours of the recent {\it Planck} results.\footnote{We would like to thank
 Laila Alabidi for providing us with the data for these confidence contours.}

When we allow for a non-zero $f_{\chi\chi}$, we see that it tends to
make the spectrum more blue-tilted, whilst leaving the tensor-to-scalar
ratio unaffected.  This again can be easily understood.  When we
consider the contribution to the spectral tilt from the Riemann curvature term,
the only non-zero component is proportional to
$R^\chi{}_{\phi\chi\phi}$.  Namely, we have the additional term
\begin{equation}\label{eq:19}
 n_s-1\supset \frac{2\mathcal{N}_\chi^2S^{\chi\chi}R^\chi{}_{\phi\chi\phi}\left(\frac{d\phi}{d\tilde{t}}\right)^2}{3\tilde{H}^2\mathcal{N}_a\mathcal{N}_bS^{ab}}.
\end{equation}
For the values of $f_\chi\lesssim \mathcal{O}(0.1)$ that we are
considering, we find that to leading order $R^\chi{}_{\phi\chi\phi}\sim
f_{\chi\chi}/(2f)$.  Given that all the other factors
in \eqref{eq:19} are positive definite, we therefore see that a positive
non-zero $f_{\chi\chi}$ will give a positive contribution to $n_s-1$,
which is the opposite effect to a non-zero $f_\chi$. 
In the case that $p=1$, we see that the combination of a
non-zero $f_\chi$ and $f_{\chi\chi}$ can be used to bring predictions of the model
within the 68\% confidence contours of the recent {\it Planck} results.    

%\subsubsection{
\subsection{Non-gaussianity}

Let us now consider the non-gaussianity.  Given
that we are considering $f_\chi/\sqrt{f}\lesssim\mathcal{O}(\epsilon^{1/2})$ and that
$S_{ab}$ has no off-diagonal components, 
from \eqref{eq:1} and \eqref{specE}
the leading contributions to
$\tilde{f}_{NL}$ and $f_{NL}$ are given as
\begin{align}\label{eq:20}
 \frac{6}{5}\tilde{f}_{NL}&\simeq \frac{(S^{\chi\chi})^2\tilde{N}_\chi\tilde{N}_\chi\tilde{N}_{\chi\chi}}{\left[2f\tilde{N}_\phi\tilde{N}_\phi
+
S^{\chi\chi}\tilde{N}_\chi\tilde{N}_\chi\right]^2}\simeq-\frac{4\frac{f_\chi^2}{f}f_{\chi\chi}N^3}{\left[\frac{N_\ast}{p}+2\frac{f_\chi^2}{f}N^2\right]^2},\\
\label{eq:21}
 \frac{6}{5}f_{NL}&\simeq \frac{(S^{\chi\chi})^2\mathcal{N}_\chi\mathcal{N}_\chi\mathcal{N}_{\chi\chi}}{\left[2f\mathcal{N}_\phi\mathcal{N}_\phi
+
S^{\chi\chi}\mathcal{N}_\chi\mathcal{N}_\chi\right]^2}\simeq-\frac{4f_{\chi\chi}\frac{f_\chi^2}{f}\left(\frac{N(1+p)-N_\ast}{p}\right)^3}{\left[\frac{N_\ast}{p}+2\frac{f_\chi^2}{f}\left(\frac{N(1+p)-N_\ast}{p}\right)^2\right]^2},
\end{align}      
where we have also assumed $N_\ast\gg p$.  Taking
$N\sim\mathcal{O}(\epsilon^{-1})$ we see that both of these quantities are
$\mathcal{O}(1)\times f_{\chi\chi}$, so that we might expect a sizeable
$\tilde{f}_{NL}$ and $f_{NL}$ for a large $f_{\chi\chi}$.  What we must
recall, however, is that expression \eqref{eq:23} for the second
derivatives of $N$ (which we have used here) is only valid when the
curvature of the Einstein frame field-space can be neglected, and one of
the conditions we derived for this to hold was $f_{\chi\chi}\ll 1$.  As
such, at least within the confines of our analytic formulation, we find
that $\tilde{f}_{NL}$ and $f_{NL}$ will be small.  

\begin{figure}
\centering
\begin{minipage}{.5\textwidth}
  \centering
  \includegraphics[width=\linewidth]{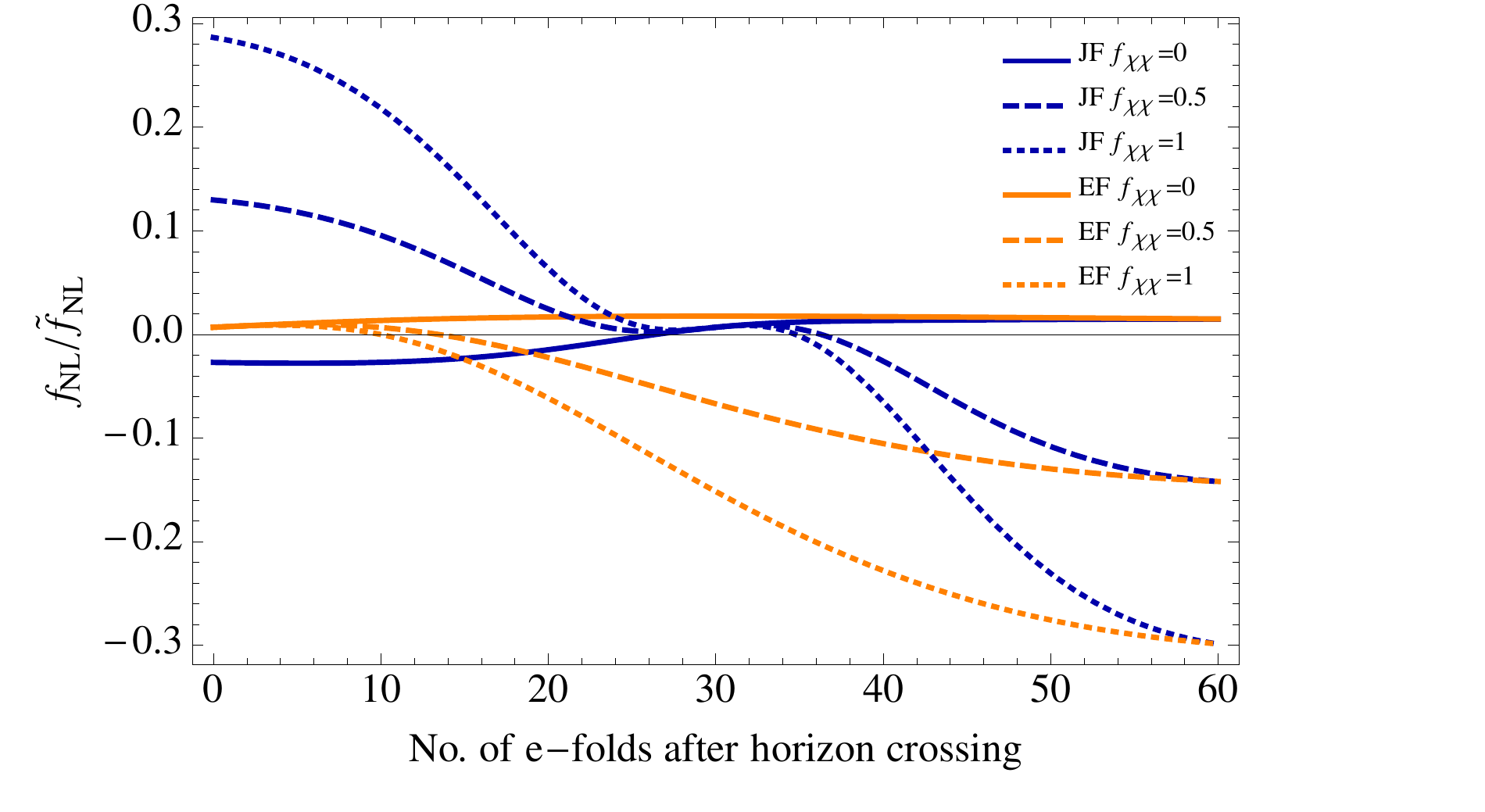}
  %\caption*{(a)}
  %\label{fig:test1}
\end{minipage}%
\begin{minipage}{.5\textwidth}
  \centering
  \includegraphics[width=.9\linewidth]{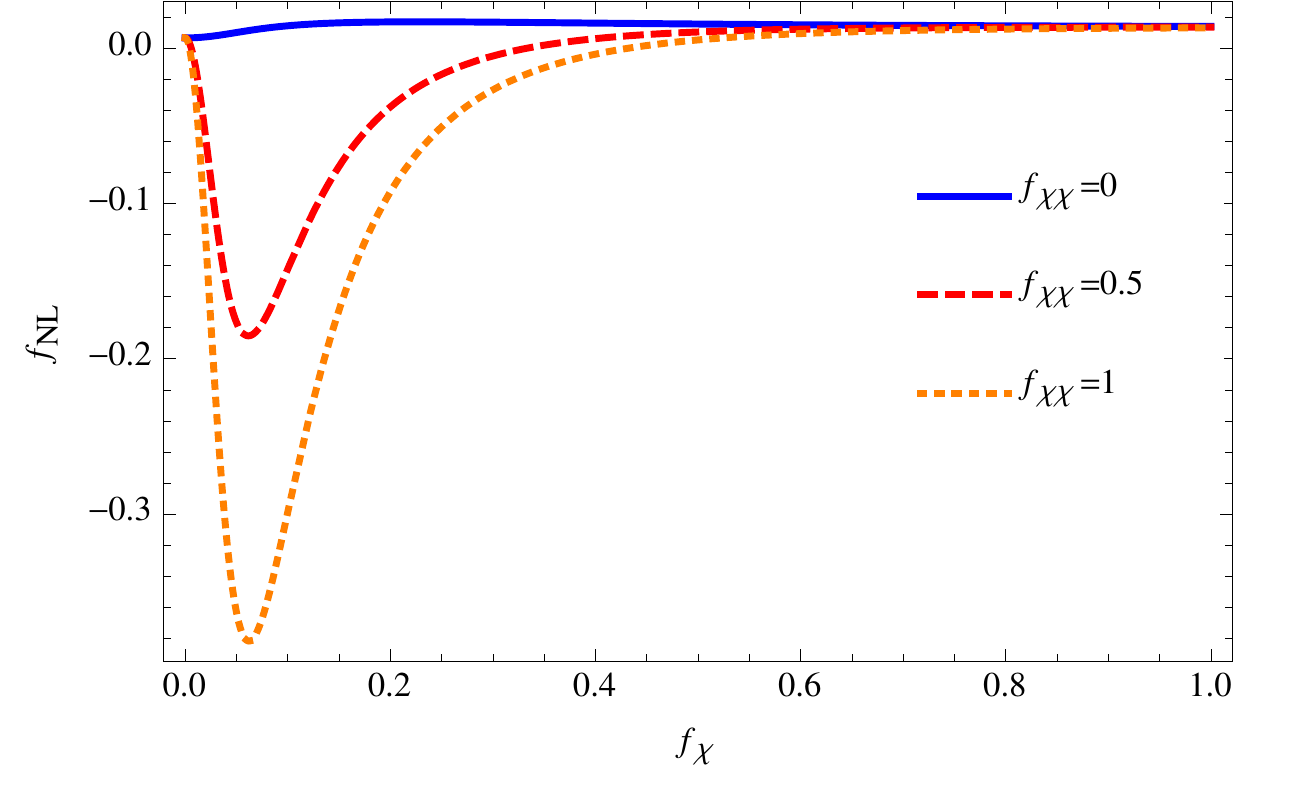}
  %\caption*{(b)}
  %\label{fig:test2}
\end{minipage}%
%\begin{minipage}{.33\textwidth}
%  \centering
%  \includegraphics[width=.9\linewidth]{fNL.pdf}
%  \caption*{(c)}
%  %\label{fig:test2}
%\end{minipage}
\caption{Left: Evolution of the $f_{NL}$ parameters in the Jordan (JF) and
 Einstein (EF) frames for a range of $f_{\chi\chi}$.  $f_\chi =
 0.1$, and the remaining
 parameters are as given in the caption of figure \ref{fig:1}.  Right:
 Dependence of the final $f_{NL}$ parameters on $f_\chi$ for a range
 of $f_{\chi\chi}$.}
\label{fig:3}
\end{figure}

In the left-hand
panel of figure \ref{fig:3} we plot the evolution of $\tilde{f}_{NL}$
and $f_{NL}$ for values of $f_{\chi\chi}$ that are at the upper limit for 
the validity of \eqref{eq:23}, with the remaining model parameters set as $2f=V^{(\chi)}=1$,
$f_\chi=0.1$, $W_{\chi\chi}=0$ and $N_\ast = 60$.  We do indeed find
that $\tilde{f}_{NL}$ and $f_{NL}$ are small.  Their evolutions, which are very different, can be
understood by referring to \eqref{eq:20} and \eqref{eq:21}.  In the
Einstein frame we see that $\tilde{f}_{NL}$ vanishes at $N=0$, but that
for $N>0$ it develops a negative value that depends linearly on
$f_{\chi\chi}$.  In the Jordan frame we see that for $N=N_\ast$ the
expression for $f_{NL}$ is the same as that for $\tilde{f}_{NL}$, and
that it is
also anti-symmetric about the time $N=N_\ast/(1+p)$, which corresponds
to $N=30$ in our case.  For $p=1$ we therefore have $f_{NL}(N=30)\simeq
0$ and
$f_{NL}(N=0)=-f_{NL}(N=N_\ast)=-\tilde{f}_{NL}(N=N_\ast)$.  

In the
right-hand panel of figure \ref{fig:3} we plot the final Jordan frame $f_{NL}$
as a function of $f_\chi$ for the same values of $f_{\chi\chi}$ as used
in the left-hand plot.  Referring to \eqref{eq:21}, it is clear that $f_{NL}\rightarrow 0$ in the
limit $f_\chi\rightarrow 0$.  Similarly, for $f_\chi^2/f\gg 1/(2N_\ast
p)$ we have $|f_{NL}|\ll 2p f_{\chi\chi}$.  There is thus some optimum
value of $f_\chi$ between these two limits where the magnitude of
$f_{NL}$ is peaked, and from the figure \ref{fig:3} we see that this value is
between $0.05$ and $0.1$.  However, even the peak magnitude $f_{NL}\sim \mathcal{O}(0.1)$
is very small, which is consistent with the {\it Planck} constraint $f_{NL}=2.7\pm 5.8$.    

\section{Non-minimally coupled multi-brid inflation}\label{nmmbsec}

In this section, as a second example, we extend the multi-brid inflation model introduced
in \cite{multibrid} to include non-minimal coupling. 

\subsection{The model}

We consider
the potential and non-minimal coupling to take the form
\begin{equation}
\label{mb_pot}
V = V_0\exp\left[\sum_am_a\phi^a\right]\qquad\mbox{and}\qquad f =
\sqrt{f_0}\exp\left[\sum_a\frac{z_a}{2}\phi^a\right],
\end{equation}
such that the effective potential in the Einstein frame is of the same form as that 
in the original model \cite{multibrid} and is product separable, namely 
\begin{equation}
\label{mb_coup}
W = \frac{V_0}{f_0}\exp\left[\sum_a(m_a - z_a)\phi^a\right],
\end{equation}   
where $m_a$ and $z_a$ are constants.  The details of the model are
presented in appendix \ref{AppNMM}, and here we simply discuss the key
results.  In some sense the multi-brid model is very similar to that
considered in the previous section, as the trajectory in field-space
is once again straight, meaning that the decomposition of the fields
into adiabatic and isocurvature components with respect to the Jordan
frame metric is time independent.  Unlike in the previous model,
however, here we allow both fields to be non-minimally coupled.

As in the original model \cite{multibrid}, we assume that inflation is
terminated via the instability of a hybrid inflation-type waterfall
field $\chi$, which is implemented by replacing $V_0$ with
\begin{equation}
V_0 = \frac{1}{2}\sum_aw_a^2(\phi^{a})^2\chi^2 +
\frac{\lambda}{4}\left(\chi^2-\frac{\sigma^2}{\lambda}\right)^2,
\end{equation}
such that the field $\chi$ develops a tachyonic mass for
$\sum_aw_a^2(\phi^a)^2<\sigma^2$.  Taking a two-field example, we can
thus parameterise the field values at the end of inflation as
$\phi^1_\diamond = \sigma \cos \gamma/w_1$ and $\phi^2_\diamond = \sigma
\sin \gamma/w_2$.

\subsection{%Contribution from after the end of inflation
Computing $\delta N$}

Using the fact that 
$$
\frac{\phi^1}{m_1-z_1}-\frac{\phi^2}{m_2-z_2}={\rm constant}, 
$$ 
one can determine how $\gamma$ is related to the initial field values $\phi_\ast^1$ and
$\phi^2_\ast$.  Perturbing this relation one can subsequently find $\delta\gamma$ in terms of the initial field fluctuations $\delta\phi_\ast^1$ and
$\delta\phi_\ast^2$, which in turn allows us to expand $\delta N$ in
terms of the field perturbations on the initial flat hypersurface.  
Expanding $\delta N^\sigma$ up to second order, where we have added the
superscript $\sigma$ to denote the fact that this is $\delta N$ up to
the surface in field space defined by the waterfall transition, we find
\begin{align}\nonumber
 \delta N^\sigma &=
\frac{1}{2f_\diamond}\frac{w_2\sin\gamma\delta\phi^2_\ast+w_1\cos\gamma\delta\phi^1_\ast}{M_2w_2\sin\gamma+M_1w_1\cos\gamma}+\frac{1}{2f_\diamond}\frac{(w_1
w_2)^2}{2\sigma}\frac{(M_1\delta\phi^2_\ast-M_2\delta\phi^1_\ast)^2}{(M_2w_2\sin\gamma
+ M_1w_1\cos\gamma)^3}\\
&\quad-\frac{N_\ast}{2}(z_1\delta\phi^1_\ast
+ z_2\delta\phi^2_\ast)+\frac{N_\ast}{8}( z_1\delta\phi^1_\ast +
z_2\delta\phi^2)^2-\frac{\delta\mathcal{S}}{8f_\diamond(M_2w_1\cos\gamma +
M_2w_2\sin\gamma)^2},\label{eq:13}
\end{align}
where $M_a=m_a-z_a$ are the effective masses of the fields and $\delta\mathcal{S}$ is given as
\begin{align}\nonumber
\delta\mathcal{S}&=w_1\cos\gamma(\delta\phi^1_\ast)^2(w_1\cos\gamma(z_1M_1-z_2M_2)+2z_1M_2w_2\sin\gamma)\\\nonumber
&\quad+
2\delta\phi^1_\ast\delta\phi^2_\ast(z_2M_1w_1^2\cos^2\gamma+z_1M_2w_2^2\sin^2\gamma)\\
&\quad+w_2\sin\gamma(\delta\phi^2_\ast)^2(2z_2M_1w_1\cos\gamma-w_2\sin\gamma(z_1M_1-z_2M_2)).
\end{align}  
From \eqref{eq:13} it is easy to see that in the limit
$z_1,z_2\rightarrow 0$, which also gives $2f_\diamond\rightarrow 1$ and
$M_a\rightarrow m_a$, the second line vanishes and we recover the result
of \cite{multibrid}.  As such, we see that the main effect of
introducing non-minimal coupling, aside from $m_a\rightarrow M_a$, is
the appearance of the additional terms on the second line of
\eqref{eq:13}.  To determine the importance of these additional terms,
we note that in order to satisfy the slow-roll conditions \eqref{eq:14}-\eqref{sr3}
we require $z_a,M_a\sim\mathcal{O}(\epsilon^{1/2})$ and
$z_a\phi^a\lesssim\mathcal{O}(1)$, where we are assuming
$\epsilon\sim\eta^{(a)}\sim\delta$.  Taking
$N_\ast\sim\mathcal{O}(1/\epsilon)$, we therefore find that the
additional linear terms on the second line of \eqref{eq:13} are of the
same order of magnitude as those on the first line.  Furthermore, the
explicit dependence on $N_\ast$ would suggest that these terms will
give rise to a significant scale dependence of the spectrum.  Turning to the
additional second order terms, we find that they will generally be suppressed by
$\mathcal{O}(\sigma\epsilon^{1/2}/w)$ relative to those on the first
line, where we have assumed $w_1\sim w_2=w$.
This would suggest that the non-minimal coupling does not induce
substantial additional non-gaussianity in comparison to the minimally
coupled case.

It is important to recall that when applying the $\delta N$ formalism we
must take the final surface up to which $N$ is calculated to be one of
constant energy density.  As the end-of-inflation surface defined above
in terms of the waterfall transition does not generally coincide with
one of constant energy, we must therefore determine the
additional contribution to $\delta N$ coming from evolution between the
end-of-inflation surface and one of constant energy during the radiation
dominated era that follows.  Namely, we have  
\begin{equation}
 \delta N = \delta N^\sigma + \delta N^r,
\end{equation}   
where the superscript $r$ denotes the contribution just described.
Interestingly, consideration of this additional contribution brings to
light a subtlety regarding matching conditions in the Jordan and
Einstein frame analyses.  Referring back to \eqref{eq:15}, and noting
that the terms in the second line will vanish if we assume that in the
radiation domination era we have $2f = 1$ independent of the initial
field values, the difference between the curvature perturbations in the
two frames will be given as 
\begin{equation}
 \zeta-\tilde{\zeta}=\left[N_a^\sigma + N_a^r - N_a^{\tilde{\sigma}} -
		      N_a^{\tilde{r}}\right]\delta\phi^a_{\tilde{\mathcal{R}}}+\frac{1}{2}\left(N_{ab}^\sigma+N_{ab}^r-N_{ab}^{\tilde{\sigma}}-N_{ab}^{\tilde{r}}\right)\delta\phi^a_{\tilde{\mathcal{R}}}\delta\phi^b_{\tilde{\mathcal{R}}}. 
\end{equation}
Seeing as the waterfall instability condition for the end of inflation
$\sum_aw_a^2(\phi^a)^2=\sigma^2$ is independent of the frame we have
$N^\sigma_a = N^{\tilde{\sigma}}_a$, and similarly for the second
derivatives.  As such, any difference between $\zeta$ and $\tilde{\zeta}$
is determined by the relation between $N^r$ and $N^{\tilde{r}}$.   

In determining $N^r$ and $N^{\tilde{r}}$ we assume instant reheating.  At
background level, the Israel Junction conditions then tell us that the
scale factor and Hubble rate should be continuous across the matching
surface (see e.g. \cite{ArrojaWands}).  However, here we have the scale factors and Hubble rates as
defined in the Jordan and Einstein frames, and so it seems we must
decide which we would like to be continuous, i.e. we must constrain
either
\begin{equation}\label{matchingchoice}
[a]^+_- = 0\quad\mbox{and}\quad[\mathcal{H}]^+_-=0\qquad\mbox{or}\qquad [\tilde{a}]^+_- = 0\quad\mbox{and}\quad[\tilde{\mathcal{H}}]^+_-=0,  
\end{equation}  
where $[a]^+_- = a_+-a_-$ and $+/-$ label values immediately after/before the matching surface.   
If we require that the Einstein frame quantities be continuous, then we find that 
\begin{equation}
a_+=\sqrt{2f_-}a_-= \tilde{a}_+\qquad\mbox{and}\qquad \mathcal{H}_+ = \mathcal{H}_- + \frac{f^\prime_-}{2f_-} = \tilde{\mathcal{H}}_+,
\end{equation}
or if we require that the Jordan frame quantities be continuous we find
\begin{equation}
\tilde{a}_+=\frac{\tilde{a}_-}{\sqrt{2f_-}}= a_+\qquad\mbox{and}\qquad \tilde{\mathcal{H}}_+ = \tilde{\mathcal{H}}_- - \frac{f^\prime_-}{2f_-} = \mathcal{H}_+,
\end{equation}
where in both cases we take $2f_+ =1$ so that the Jordan and Einstein
frames coincide in the radiation dominated phase, after all fields
have decayed.  Due to the discontinuity in $f$, we see that enforcing
continuity in one frame leads to a discontinuity in the other.  In
particular, the energy density (given by $\rho=3H^2$ and $\tilde{\rho}=3\tilde{H}^2$ in the Jordan and Einstein frames respectively) will
be discontinuous in one of the frames, in order that finally $\rho_+=\tilde{\rho}_+$. 

Making the instant reheating approximation, and using the fact that $\rho_+=\tilde{\rho}_+$, the number of e-foldings in the radiation dominated phase is given as 
\begin{equation}
N^{r} = \frac{1}{4}\ln \left(\frac{\rho_{+}}{\rho_r}\right)=\frac{1}{4}\ln \left(\frac{\tilde{\rho}_{+}}{\rho_r}\right)=N^{\tilde{r}}%=\frac{1}{4}\ln \left(\frac{3H^2_{+}}{\rho_r}\right)=\frac{1}{4}\ln \left(\frac{3\tilde{H}^2_{+}}{\rho_r}\right),
\end{equation} 
where $\rho_r$ corresponds to some final constant density in the radiation
dominated phase.  
%Once we have chosen in which frame to apply the
%matching conditions, the fact that $\rho_+=\tilde{\rho}_+$ means that
%$N^r=N^{\tilde{r}}$.  
%However, 
%we also need to include the contribution of the end-of-inflation surface
%into the number of e-fold.
%Then, 
Although we do have $N^r=N^{\tilde{r}}$, their value will depend on whether we
choose to require continuity of the scale factor and Hubble rate in the
Jordan or Einstein frame.  Namely, we need to specify whether $\rho_{+}
= \rho_{-}$ or $\rho_{+} = \tilde{\rho}_{-}$.  
%If, 
Requiring continuity of the scale factor and Hubble rate in the Jordan and Einstein frames
we find
%the number of e-folding in the radiation dominated phase including the contribution of
%the end-of-inflation surface is given by 
\begin{equation}
N^r_{\pm} = N^{\tilde{r}}_{\pm} = \frac{1}{4}\ln\left(\frac{\rho_-}{\rho_r}\right)\quad\mbox{and}\quad N^r_{\tilde{\pm}} = N^{\tilde{r}}_{\tilde{\pm}} = \frac{1}{4}\ln\left(\frac{\tilde{\rho}_-}{\rho_r}\right),
\end{equation}    
respectively,
where $\pm$ and $\tilde{\pm}$ denote the continuity conditions being
applied in the Jordan and Einstein frames.  
For simplicity, if we
assume that the slow-roll approximation is still valid at the end of
inflation, then we have that $\tilde{H}_- = H_-/\sqrt{2f_-}$, i.e. $\tilde{\rho}_- = \rho_-/2f_-$,
from which we deduce
\begin{equation}
 N^r_{\pm} - N_{\tilde{\pm}}^{r} = \frac{1}{4}\ln(2f_-).
\end{equation}    

Calculating the terms explicitly, we find
\begin{align}\nonumber
\delta N^r_{\pm} &=
\frac{1}{4}\left\{\left(m_2-\frac{z_2}{2}\right)\frac{\sigma\cos\gamma}{w_2}-\left(m_1-\frac{z_1}{2}\right)\frac{\sigma\sin\gamma}{w_1}\right\}\left(\delta\gamma_{(1)}+\delta\gamma_{(2)}\right)\\
&\quad-\frac{1}{4}\left\{\left(m_2-\frac{z_2}{2}\right)\frac{\sigma\sin\gamma}{w_2}+\left(m_1-\frac{z_1}{2}\right)\frac{\sigma\cos\gamma}{w_1}\right\}\left(\delta\gamma_{(1)}\right)^2,\\
\delta N^r_{\pm} - \delta N^{r}_{\tilde{\pm}}&=
\frac{1}{4}\left\{\frac{z_2}{2}\frac{\sigma\cos\gamma}{w_2}-\frac{z_1}{2}\frac{\sigma\sin\gamma}{w_1}\right\}\left(\delta\gamma_{(1)}+\delta\gamma_{(2)}\right)-\frac{1}{4}\left\{\frac{z_2}{2}\frac{\sigma\sin\gamma}{w_2}+\frac{z_1}{2}\frac{\sigma\cos\gamma}{w_1}\right\}\left(\delta\gamma_{(1)}\right)^2.
\end{align} 
Comparing the coefficients in these expressions with those in
(\ref{mulhyb}-\ref{mulhybend}), we see that they are suppressed by a
factor $\sim \mathcal{O}(\epsilon)$, where we have used the fact that
$m_a,z_a\sim\mathcal{O}(\epsilon^{1/2})$ in order that the slow-roll
conditions be satisfied.  As such, we find that the contribution of
$\delta N^r$ to $\delta N$ is subdominant, as in the minimally coupled case
\cite{multibrid}.  Ultimately, by considering in more detail the mechanism by which reheating
takes place in this type of model, one would hope to be able to determine
which of the matching conditions in \eqref{matchingchoice} is appropriate in the 
instant reheating limit.  %Such work is currently under way.

\subsection{%An explicit example based on
Comparison with the original %multi-brid 
model}

In order to determine more explicitly the effect of the non-minimal
coupling, let us consider an example.  We focus on the parameter
space evaluated in the original %multi-brid 
paper \cite{multibrid}, taking
$M_1^2=0.005$, $M_2^2=0.035$, $\gamma\rightarrow 0$ and $w_1 = w_2 = w$.  We
then further choose $w=0.1$, $\sigma = 10^{-2}$, $2\sqrt{f_0}=1$ and
$N_\ast = 60$.
In the case that $z_1=z_2=0$, these parameters give $n_s = 0.96$,
$r=0.04$, $f_{NL}\sim 4.1$ and also satisfy the condition $\sigma\gg
H$, 
which is required in order that the $\chi$ field can act as a waterfall field
and support a sharp phase transition.
%which is required in order that the waterfall field mass be large
%enough during inflation. 
 Having established that $\delta N^{r}\ll\delta
N^\sigma$, we focus only on $\delta N^\sigma$, given in
\eqref{eq:13}.  In this example the general expression simplifies to 
\begin{align}\label{eq:16}
 \delta N^\sigma &\simeq
\frac{1}{2f_\diamond}\frac{\delta\phi^1_\ast}{M_1}-\frac{N_\ast}{2}(z_1\delta\phi^1_\ast+z_2\delta\phi^2_\ast)+\frac{1}{2f_\diamond}\frac{w}{2\sigma}\frac{(M_1\delta\phi^2_\ast-M_2\delta\phi^1_\ast)^2}{M_1^3},
\end{align}
where we have omitted the additional second-order terms on account of
their being small.  When using the coefficients of this expression in
the general formulae for the curvature power spectrum, spectral tilt and $f_{NL}$
parameter, we make the additional simplification $S_{ab}\sim
\delta_{ab}/2f$ and also neglect the curvature term in the expression
for the spectral tilt.  We know that these approximations are valid due to the
fact that $z_a\sim \mathcal{O}(\epsilon^{1/2})$.  

\begin{figure}
\centering
\begin{minipage}{.33\textwidth}
  \centering
  \includegraphics[width=\linewidth]{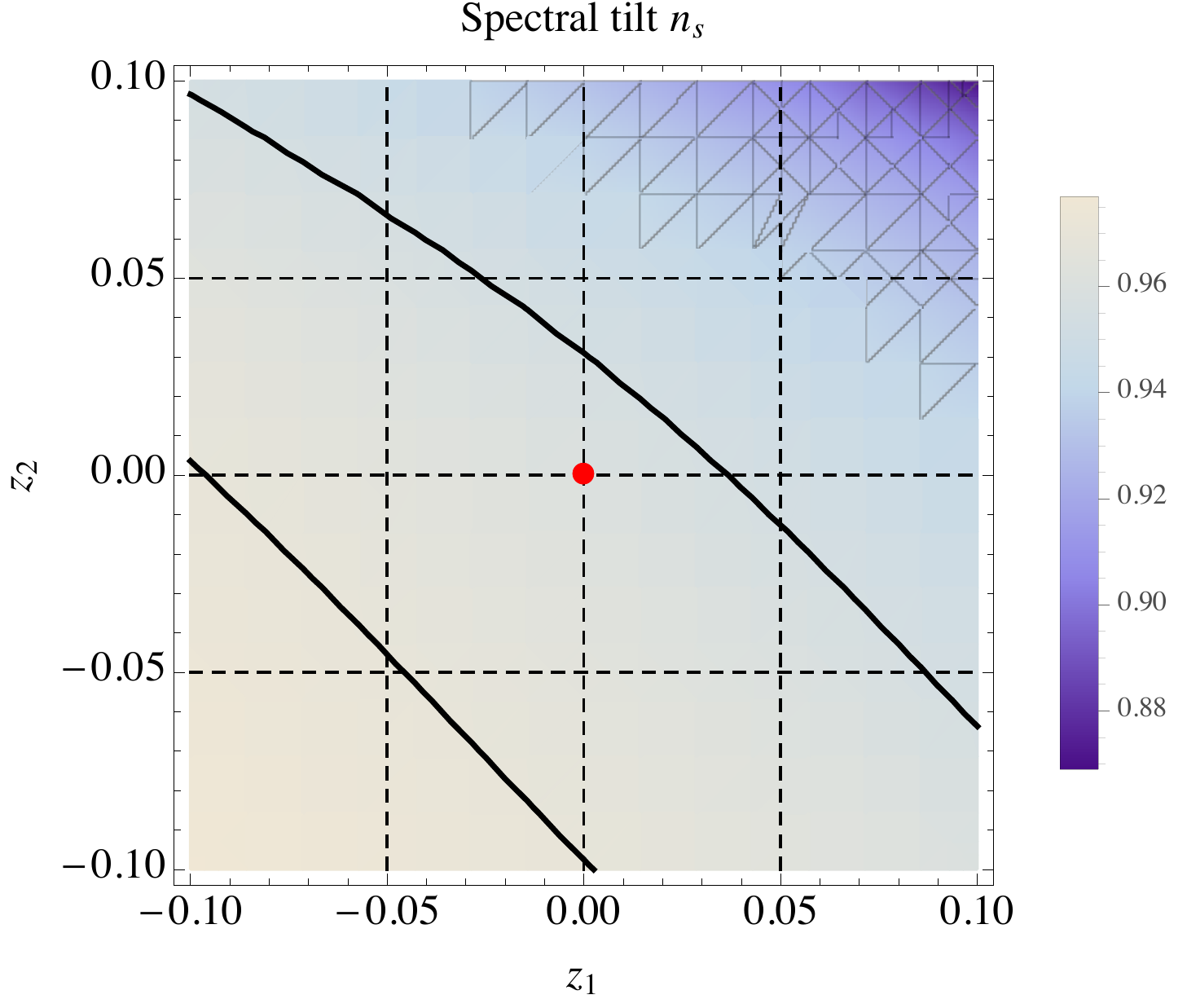}
  %\caption*{(a)}
  %\label{fig:test1}
\end{minipage}%
\begin{minipage}{.33\textwidth}
  \centering
  \includegraphics[width=\linewidth]{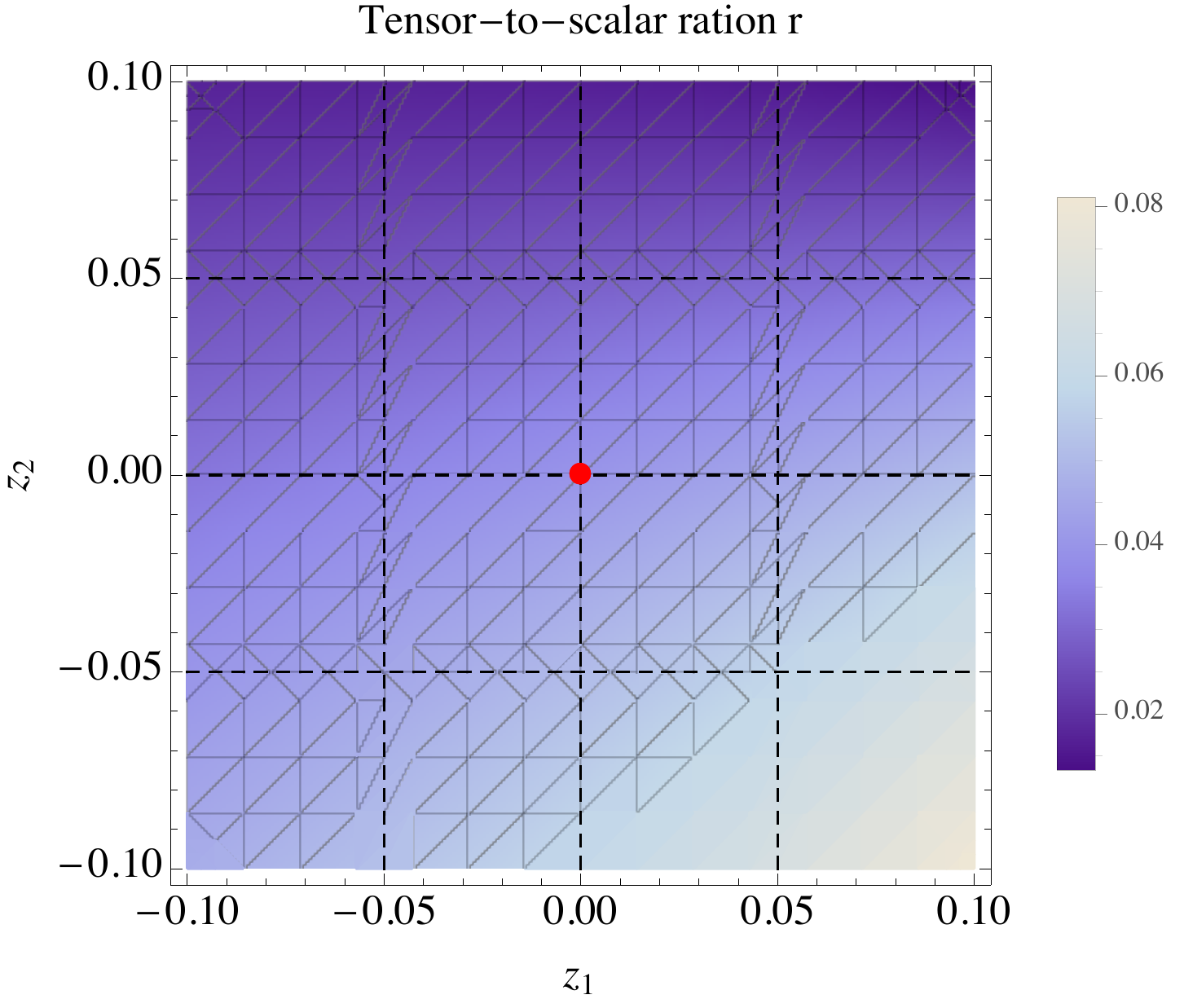}
  %\caption*{(b)}
  %\label{fig:test2}
\end{minipage}
\begin{minipage}{.33\textwidth}
  \centering
  \includegraphics[width=\linewidth]{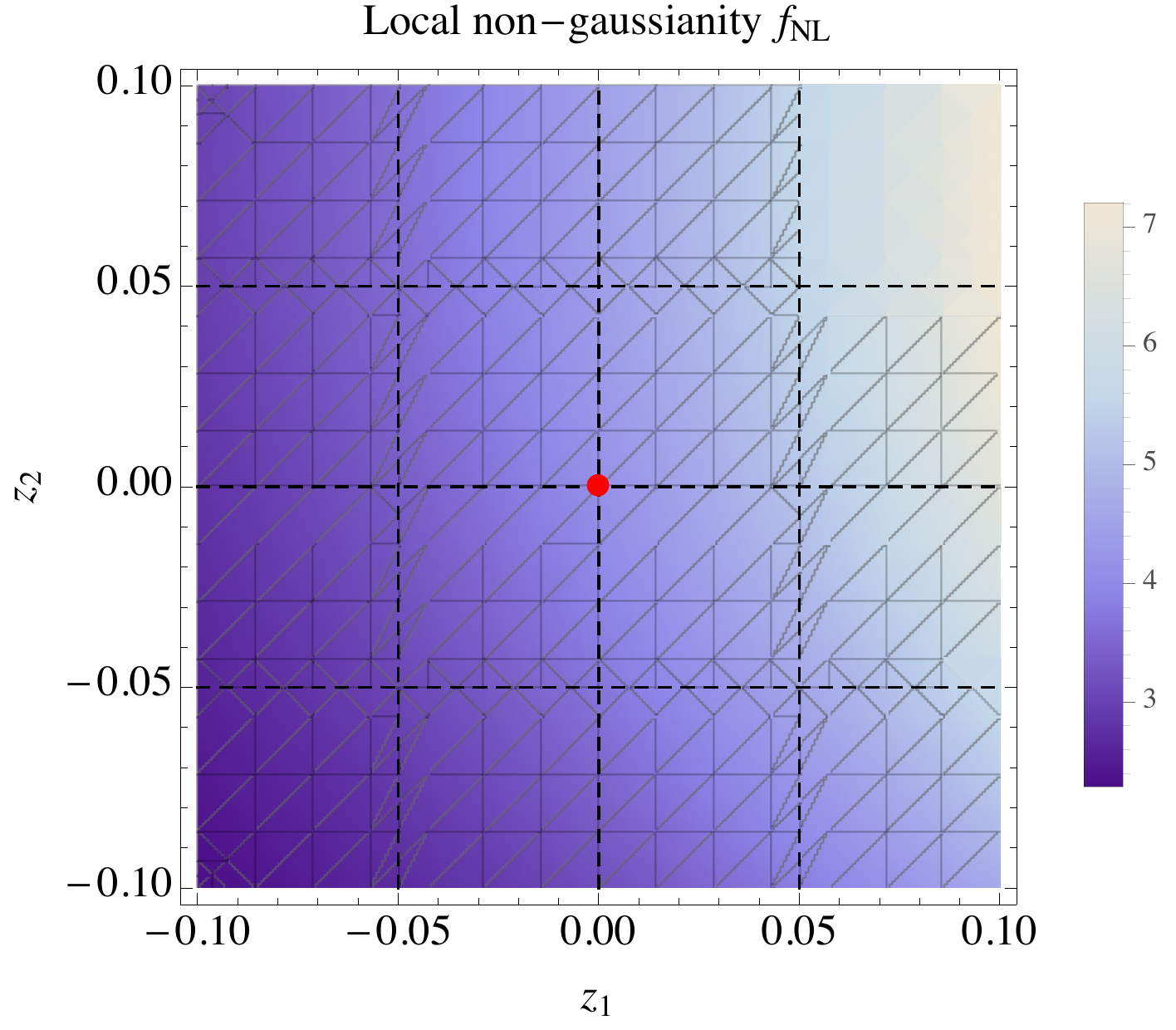}
  %\caption*{(c)}
  %\label{fig:test2}
\end{minipage}
\caption{\small{Dependence on $z_1$ and $z_2$ of the spectral tilt (left), tensor-to-scalar ratio
 (centre) and $f_{NL}$ parameter (right) in the multi-brid inflation
 model with non-minimal coupling.  The red point at $(z_1,z_2)=(0,0)$
 corresponds to the minimally coupled case considered in
 \cite{multibrid}.  The model parameters are $M_1^2=0.005$,
 $M_2^2=0.035$, $\gamma\rightarrow 0$, $w_1 = w_2 = w =0.1$, $\sigma = 10^{-2}$, $2\sqrt{f_0}=1$ and
$N_\ast = 60$, which for $z_1=z_2=0$ give $n_s= 0.96$,
$r=0.04$ and $f_{NL}\sim 4.1$.  The black curves shown on the
 left-hand plot correspond to the constraint $n_s=0.9697\pm 0.0073$
 coming from the recent {\it Planck} results.}}  
\label{fig:nmm}
\end{figure}

Plots of the power spectrum, spectral tilt and $f_{NL}$ parameter as functions of
$z_1$ and $z_2$ are shown in figure \ref{fig:nmm}.  For all the plotted
values of $z_1$ and $z_2$, we find that $r$ and $f_{NL}$ are consistent
with the recent {\it Planck} results $r<0.11$ and $f_{NL}=2.7\pm 5.8$.
In the plot of $n_s$ we include curves corresponding to the 68\% confidence interval
obtained by the {\it Planck} collaboration, namely $n_s=0.9697\pm 0.0073$.  In all the
plots, the central red point at $(z_1,\,z_2)=(0,\, 0)$ corresponds
to the minimally coupled case considered in \cite{multibrid}.  The
dependencies of the three parameters on $z_1$ and $z_2$ are relatively
easily understood.  

First, regarding the spectral tilt, under the assumptions outlined above we find
\begin{equation}
 n_s-1=-2f_\ast(M_1^2+M_2^2)+2f_\ast(\frac{z_1}{2}M_1+\frac{z_2}{2}M_2)-\frac{z_1N_1+z_2N_2}{N_1^2+N_2^2},
\end{equation}
where $N_1$ and $N_2$ are the coefficients of $\delta\phi^1_\ast$ and
$\delta\phi^2_\ast$ in \eqref{eq:16} respectively.  For our choice of
parameters, we find that the dominant term is
always the
the first one, and its magnitude will increase with $z_1$ and $z_2$ as a
result of the exponential growth of $f_\ast$ with $z_1$ and $z_2$.  Correspondingly, the spectrum becomes
more red-tilted.  

Next, turning to the tensor-to-scalar ratio, we have
\begin{equation}
 r = \frac{8}{2f_\ast\left[N_1^2+N_2^2\right]}.
\end{equation} 
Considering the $z_1$-dependence in the quadrant where $z_1,z_2>0$, we find that
there are two competing effects as we increase $z_1$: $N^2_1$ will get
smaller whilst $f_\ast$ will increase.  Which effect ``wins'' depends on
the value of $z_2$, and we find that for most values of $z_2$ the first
effect wins, meaning that $r$ increases with $z_1$.  Considering the
dependence on $z_2$, both $N_2^2$ and $f_\ast$ increase with $z_2$,
leading to a decrease in $r$.  

Turning now to $f_{NL}$, and neglecting the subdominant terms associated
with Christoffel symbols such that $\tilde{\nabla}_a\tilde{\nabla}_bN
\rightarrow N_{ab}$, we find that the factors
of $2f_\ast$ coming from $S_{ab}$ cancel and we are simply left with
\begin{equation}
\frac{6}{5}f_{NL}=\frac{N_1^2N_{11}+2N_1N_2N_{12}+N_2^2N_{22}}{\left(N_1^2+N_2^2\right)^2},
\end{equation}
where $N_{11}$, $N_{12}$ and $N_{22}$ can be read off from
\eqref{eq:16}.  As the second derivatives of $N$ do not depend on $z_1$
and $z_2$, we only need consider the behaviour of first derivatives.
$N_1$ decreases with increasing $z_1$, and seeing as the denominator goes as
$N_1^4$ whilst the numerator only as $N_1^2$, we find that $f_{NL}$
increases with $z_1$.  Regarding the $z_2$ dependence, although the
$N_2^2$ term in the denominator increases, it remains subdominant
compared to $N_1^2$.  As such, the important $z_2$ dependence comes from
the $N_1N_2N_{12}$ term in the numerator.  The minus signs from $N_2$
and $N_{12}$ cancel meaning that this term causes $f_{NL}$ to increase
with $z_2$.

As a final comment, we point out that in the non-minimal extension of
the multi-brid model there is a contention between achieving enough
e-folds of inflation and satisfying the slow-roll conditions at the
beginning of inflation when we take $z_1,z_2>0$.  From expression \eqref{eq:17}
for $N$ we see that a relatively large value for $f_\ast$ is required in
order to achieve $N\sim 60$ if $z_a\sim 0.1$.  However, from
\eqref{eq:14}-\eqref{sr3}, we see that this will tend to lead to an enhancement of
the slow-roll parameters.  As an example, let us consider the largest slow
roll parameter, $\eta$.  If we were to set $z_1 = z_2 = z$, require $N =
60$ and $\eta_\ast \lesssim  0.1$, we find the condition $z\lesssim
0.06$.  Alternatively, if we consider $N=60$ and $z_1 = z_2 = 0.1$, then
this gives us $\eta_\ast \sim 0.3$.  As such, referring to figure
\ref{fig:nmm}, we see that the results obtained in the top right corner
of these figures are somewhat invalidated as a result of the breaking of slow-roll.

\section{Summary and Conclusions}\label{concsec}

Using the $\delta N$ formalism we have considered the non-linear
curvature perturbation arising in multi-field models of inflation with
non-minimal coupling.  With an analysis in the Einstein frame having
already been developed in \cite{eliston}, here we have focused on the
relation between the $\delta N$ formalism as applied in the Jordan and
Einstein frames, which in turn allowed us to determine how the curvature
perturbation in the Einstein frame can be related back to that in the
Jordan frame.  Key in our analysis was finding the relation between the
flat-gauge field perturbations as defined in the Jordan and Einstein
frames, $\delta\phi^a_{\mathcal{R}}$ and
$\delta\phi^a_{\tilde{\mathcal{R}}}$ respectively.  Using this relation,
in combination with the relation between $N$ and $\tilde{N}$, we were
able to show that the difference in definition of the initial flat
hypersurface associated with the $\delta N$ analysis in the two frames
does not give rise to a difference in the final curvature perturbations.
As such, it is only the difference in definition of the final constant
energy hypersurface that leads to $\zeta\ne\tilde{\zeta}$.  This result
seems to agree with what one might expect.  In the case
that an adiabatic limit is reached, where the final constant energy
surface is uniquely defined, we therefore recover the expected result
$\zeta=\tilde{\zeta}$.  Relating $\delta\phi^a_\mathcal{R}$ to
$\delta\phi^a_{\tilde{\mathcal{R}}}$ also allowed us to exploit the
known correlation functions of $\delta\phi^a_{\tilde{\mathcal{R}}}$ to
determine the power spectrum, spectral tilt and $f_{NL}$ parameter associated with
the Jordan frame curvature perturbation.

Having discussed the general formalism in section
\ref{generalformalism}, we then considered the class of analytically
soluble, slow-roll inflation models where the Einstein frame potential is product
separable, eventually focusing on two specific examples.  The first
example consisted of a minimally coupled inflaton field $\phi$ and a
non-minimally coupled `spectator' field $\chi$.  We considered a
potential of the form $V^{(\phi)}=m^2\phi^{2p}$ for the inflaton, and also
took the $\chi$ field to be massless.  We found that the curvature
perturbations in the Jordan and Einstein frames evolve very differently
after horizon crossing, as do the corresponding power spectra, spectral tilts and
$f_{NL}$ parameters, but that by the end of inflation the two quantities
converge to leading order in the slow-roll approximation.  In terms of
the final predictions for the curvature perturbation, for $p=0.5$ and
$p=1$ we found it possible to bring predictions for the observables $r$ and
$n_s$ well within the 68\% confidence contours of the
recent {\it Planck} results.  In particular, the presence of the $\chi$
field tends to act so as to reduce the tensor-to-scalar ratio, as well
as allowing for a wide range of spectral tilts through the choice of $f_\chi$ and
$f_{\chi\chi}$.  This can be compared to the usual spectator
field scenarios, where a negative mass for the spectator field or $p\ge
2$ is required in order to reproduce the observed red-tilted spectrum
\cite{spectator}.  In the case that the Einstein frame field-space curvature is
taken to be small, the model also predicts a very small $f_{NL}$,
which again is in agreement with the recent {\it Planck} results.  We remark that  
in this model we have referred to $\chi$ as a `spectator' field in order to reflect the fact 
that it is non-dynamical at background level.  However, our model is not a spectator 
field model in the usual sense, as $\chi$ 
does contribute to the curvature perturbation throughout inflation.

The second example we considered was a non-minimally coupled extension
of the multi-brid inflation model.  As the end-of-inflation surface is
defined by the tachyonic instability of a waterfall field, which is a
frame independent definition, the Jordan and Einstein frame
curvature perturbations are identical. 
Using the same model parameters
as considered in the original multi-brid model, we found that the
introduction of non-minimal coupling has a significant effect on the
predictions of the model.  In particular, we found the ranges $0.87\lesssim n_s
\lesssim 0.98$, $0.02\lesssim r
\lesssim 0.08$ and $2\lesssim f_{NL}\lesssim 7$ for the range of
non-minimal coupling parameters considered.  These should be compared
with the results $(n_s,\,r,\,f_{NL})\sim (0.96,0.04,4.1)$ in the minimally
coupled case.  As such, we see that the {\it Planck} result $n_s =
0.9624 \pm 0.0073$ can be used to constrain the non-minimal coupling
parameters.  

As a closing remark we note that throughout this paper we have been
implicitly making the slow-roll assumption by assuming that $N$ and
$\tilde{N}$ are only functions of the initial field values and not the
initial field velocities.  Extending to the more general case
involves a number of subtleties.  For example, the horizon-crossing
conditions in the Jordan and Einstein frames, $k=aH$ and
$k=\tilde{a}\tilde{H}$ respectively, are no longer equivalent
outside of the slow-roll approximation.  
Generalisation to the non
slow-roll case is currently under consideration.

\acknowledgments

We would like to thank Joseph Elliston, Jinn-Ouk Gong and Yuki Watanabe for helpful discussions.  J.\,W. was supported by Grant-in-Aid for the Global COE Program ``The Next Generation of Physics, Spun from Universality and Emergence" from the Ministry of Education, Culture, Sports, Science and Technology (MEXT) of Japan.  M.\,M. was supported by the Yukawa fellowship and by Grant-in-Aid for
Young Scientists (B) of JSPS Research, under Contract No. 24740162.
The work of M. M. was also funded by {\it Funda\c{c}\~ao Para a Ci\^encia e a Tecnologia} (FCT)-Portugal,
through the grant SFRH/BPD/88299/2012.  This work was also supported in part by JSPS Grant-in-Aid
for Scientific Research (A) No.~21244033.

\appendix

%\appendixpage

%\addappheadtotoc

%\section{Justifying use of the $\delta N$ formalism???}
%\section{Extension to non-slow-roll case}

\section{Analytically soluble models in the Einstein frame}\label{EFappend}

In the main text we chose
to work in the Jordan frame when considering analytically soluble models.  
In this appendix we confirm that we could
equally well have started out in the Einstein frame.   

Varying the Einstein frame action \eqref{EAc} with respect to the fields $\phi^a$ and metric $\tilde{g}_{\mu\nu}$,
%In the Einstein frame, 
at background level % of the metric \eqref{metDecompE} 
we find the equations of
motion and Friedmann equation are given as
\begin{align}\label{FullEFEOM}
\frac{D^{{}^{(S)}}%\phi^{\prime a}
}{d\tilde{t}}
\Big(\frac{d\phi^a}{d\tilde t}\Big)
+3\tilde{H}%\phi^{\prime a}
\frac{d\phi^a}{d\tilde t}
+\frac{1}{4}S^{ab}W_{b}=0\quad\quad\mbox{and}\quad\quad 3\tilde{H}^2 =
\frac{1}{2}S_{ab}
%\phi^{\prime a}
%\phi^{\prime b}
\frac{d\phi^a}{d\tilde t}
\frac{d\phi^b}{d\tilde t}
 + \frac{W}{4},
\end{align}
where %here the primes denote differentiation with respect to the time in the Einstein frame 
$d\tilde{t} = \sqrt{2f}dt$ is the time in the Einstein frame
and $D^{{}^{(S)}}/d\tilde{t}$ 
is the covariant derivative associated with the field-space metric
$S_{ab}$.  Even before making the slow-roll approximation one can
relatively easily
confirm the equivalence of these equations of motion with those in the
Jordan frame by using the relations between $t$ and
$\tilde{t}$, $H$ and $\tilde{H}$ and $h_{ab}$ and $S_{ab}$.  The only trick one needs to remember is to contract
the Einstein frame equations of motion with $S_{ab}$, as this removes the inverse metric $S^{ab}$
appearing in the covariant derivative, for which we don't in general have an
expression.  

In the
slow-roll limit we would approximate the above equations as
\begin{equation}\label{eFsR}
3\tilde{H}
\frac{d\phi^a}{d\tilde t}
%\phi^{\prime a}
=-\frac{1}{4}S^{ab}W_{b}\quad\quad\mbox{and}\quad\quad3\tilde{H}^2 =
\frac{W}{4},
\end{equation}
and using $d\tilde{N} = \tilde{H}d\tilde{t}$ the first of these can be
rewritten as
\begin{equation}
\frac{d\phi^a}{d\tilde{N}} = -\frac{S^{ab}W_{b}}{W},
\end{equation}
which we can solve analytically if
\begin{equation}\label{eFsC}
\frac{S^{ab}W_{b}}{W} = \frac{\tilde{g}^{(a)}(\phi^a)}{\tilde{F}({\bm
\phi})}.
\end{equation}
In line
with the Jordan frame case, we see that \eqref{eFsC} is satisfied if
\begin{equation}\label{eFmC}
S^{ab} = \frac{1}{\tilde{G}({\bm
\phi})}\rm{diag}\left(s^{(1)}(\phi^1),\,s^{(2)}(\phi^2),\,...,\,s^{(n)}(\phi^n)\right)
\end{equation}
and either
\begin{equation}
W = \prod_aW^{(a)}(\phi^a)\quad\quad\mbox{or}\quad\quad W = \sum_a
W^{(a)}(\phi^a).
\end{equation}
Due to the form of $S_{ab}$, the condition \eqref{eFmC} cannot
be satisfied if we allow the non-minimal coupling $f$ to be a function
of more than a single field (we are assuming $h_{ab} =
G({\bm\phi})\delta_{ab}$).  This appears to be much more restrictive than the
Jordan frame case.  However, as pointed out above, things become clearer
if we contract the
first of \eqref{eFsR} with $S_{ab}$.  Expanding $S_{ab}$ and $W_a$
explicitly we have
\begin{align}\nonumber
&\frac{3\tilde{H}}{2f}\left(h_{ab}
\frac{d\phi^b}{d\tilde t}
%\phi^{\prime b} 
+
\frac{3f_a %f^\prime
}{f}
\frac{df}{d\tilde t}
\right) =
-\frac{1}{4}\left(\frac{V_a}{f^2}-\frac{2Vf_a}{f^3}\right).%\\\nonumber \Rightarrow\quad
\end{align} 
Moving the second term on the left-hand-side
to the right-hand-side and rewriting in terms of 
the slow-roll parameter $\delta$ we find 
\begin{align}
\frac{3\tilde{H}}{2f}h_{ab}%\phi^{\prime b}
\frac{d\phi^b}{d\tilde t}
 &=
-\frac{1}{4}\left(\frac{V_a}{f^2}
-\frac{2Vf_a}{f^3}\left(1%-
+\frac{9\delta}{2}\right)\right)\\\nonumber
&\simeq-\frac{1}{4}\left(\frac{V_a}{f^2} -\frac{2Vf_a}{f^3}\right) =
-\frac{1}{4}W_a.
\end{align}
%\\ \Rightarrow \quad&
Finally, using $d\tilde{N} = \tilde{H}d\tilde{t}$ gives us
\begin{align}
\frac{d\phi^a}{d\tilde{N}} =
-2f\frac{h^{ab}W_{b}}{W}.
\end{align}
These equations of motion are now of exactly the same form as
in the Jordan frame, and thus the conditions for solubility are also the
same.  Note that we have not necessarily made any
assumptions about the magnitude of derivatives of $f$ with respect to the fields.  Rather, it was
the contraction $f_a%\phi^{\prime a}
\frac{d\phi^a}{d\tilde t}
\rightarrow \delta$ that allowed us
to drop the additional terms, and for $\delta \ll 1$ we
only require that $f_a$ and $%\phi^{\prime a}
\frac{d\phi^a}{d\tilde t}$ are close to being
orthogonal.

In the preceding discussion we have shown that the form for the
slow-roll equations of motion is exactly the same in the Jordan and
Einstein frames.  As such, when we integrate these equations of motion
to find $N$ and $\tilde{N}$ we will find exactly the same result, except
for any difference in the limits of integration in the two frames.  In
particular, we have not picked up the additional log term in the
Einstein frame that we expected from \eqref{Ntilde}.  This highlights
the dangers of integrating the slow-roll equations of motion, as even
slow-roll terms, whilst instantaneously suppressed, can lead to a
non-negligible effect when integrated over the full evolution.  Indeed,
in \eqref{Ntilde} we might have naively neglected the $f^\prime/2f$ term
in comparison with $\mathcal{H}$.

\section{Slow-roll conditions and the validity of \eqref{eq:23}}\label{slowrollapp}

In this appendix we clarify under what
assumptions expression \eqref{eq:23} for the second derivatives of
$N$ with respect to the initial field values is valid.
In order to do so, let us begin by taking a closer look at the slow-roll conditions
\eqref{sR1}, \eqref{sR2} and \eqref{sR3}.\footnote{This
argument follows that given in \cite{cy}, except that here we omit
factors of $n$ corresponding to the number of fields, assuming that $n$
is not too large.}  If we make the assumption
$|h_{ab}|\sim\mathcal{O}(1)$, then the first of the slow-roll conditions
gives us $\sqrt{f}W_a/W\sim\mathcal{O}(\epsilon^{1/2})$ for
all $a$.  If we then look at the third slow-roll condition, and assume
$\delta\sim\epsilon$, then we find it to be satisfied if $f_a/\sqrt{f}\sim\mathcal{O}(\epsilon^{1/2})$.
Similarly, turning to the second slow-roll condition, and assuming
$\eta^{(a)}\sim\epsilon\sim\delta$, we find it to be satisfied if
$fW_{ab}/W\sim\mathcal{O}(\epsilon)$ and
$\sqrt{f}\,^{(h)}\Gamma^a_{bc}\sim\mathcal{O}(\epsilon^{1/2})$, where
$^{(h)}\Gamma^a_{bc}$ is the Christoffel connection associated with the
Jordan frame field-space metric $h_{ab}$.  In summary, we find that the
slow-roll equations of motion are satisfied if 
\begin{eqnarray}\label{eq:11}
 \sqrt{f}\frac{W_a}{W}\sim\mathcal{O}(\epsilon^{1/2}),\quad
 \frac{f_a}{\sqrt{f}}\sim\mathcal{O}(\epsilon^{1/2}),\quad
 \sqrt{f}\,^{(h)}\Gamma^a_{bc} \sim\mathcal{O}(\epsilon^{1/2}),\quad\frac{fW_{ab}}{W}\sim\mathcal{O}(\epsilon).
\end{eqnarray}  
If we further consider
$\ddot{f}$
\begin{equation}
 \ddot{f} = f_{ab}\dot{\phi}^b\dot{\phi}^a + f_a\ddot{\phi}^a,
\end{equation}
then one can see that the condition $\ddot{f}/(fH^2)\ll 1 $ is already
satisfied as a result of the other slow-roll conditions, provided
$f_{ab}\lesssim\mathcal{O}(1)$.  If we more stringently require
$\ddot{f}/(\dot{f}H)\ll1$, then this will be satisfied if $f_{ab}\ll 1$.
Note, however, that the set of conditions derived above are not the
necessary conditions for the slow-roll approximation to be valid.  In
the case of $\dot{f}$, for example, if we have a spectator field $\chi$
satisfying $\dot{\chi}=0$, then the third slow-roll condition
\eqref{sR3} places no limit on the value of $f_\chi$.  Also note that
requiring $\sqrt{f}W_a/W\sim\mathcal{O}(\epsilon^{1/2})$ does not
necessarily constrain derivatives of $V$ and $f$ individually.  A good
example of this is the type of potential found in the Higgs inflation model \cite{higgsinf,Cervantes}, 
where
$V\propto \phi^4$ and $f\propto \phi^2$ in the large inflaton field limit ($\phi\gg1$), 
so that the combination $W=V/f^2$ is a constant.
%\footnote{For another interesting model involving the Higgs see \cite{Cervantes}, where the Higgs field is %implemented in an induced gravity model.}

Setting these final points aside, let us now return to the question of
the validity of expression \eqref{eq:23} for the second derivatives of
$N$.  Taking second derivatives of $N$ with respect to $\phi_\ast^a$
also involves taking derivatives of the equations of motion for the
scalar fields (this corresponds to the terms such as $\partial
g^{(c)}_\ast/\partial\phi^d_{\ast}\leftrightarrow
\partial\dot{\phi}^c/\partial\phi^d_\ast$ in expression \eqref{eq:23}).
In the presence of a non-flat field-space, taking such derivatives gives
rise to terms involving the curvature of the field-space, and we must be
careful to either keep track of these terms or to justify their
neglection.  In determining $f_{NL}$ we are interested in calculating
$\tilde{\nabla}_a\tilde{\nabla}_b N$, so let us consider the quantity
$\tilde{\nabla}_a (d\phi^b/d\tilde{t})$.  From the equations of motion
in the Einstein frame, which are presented in appendix \ref{EFappend}, we find\footnote{Here we follow the procedure outlined in \cite{ns}.}
\begin{align}
 \tilde{\nabla}_b\left(S_{ac}\frac{ d\phi^c}{d\tilde{t}}\right)&=-\frac{1}{3\tilde{H}}\left[3\tilde{H}_{,b}S_{ac}\frac{d\phi^c}{d\tilde{t}}+\tilde{V}_{;ab}+
 \tilde{R}_{adbc}\frac{d\phi^c}{d\tilde{t}}\frac{d\phi^d}{d\tilde{t}}\right.\\\nonumber&\left.\quad\qquad\quad+\tilde{\nabla}_b\left(\frac{d\phi^c}{d\tilde{t}}\right)\tilde{\nabla}_c\left(S_{ad}\frac{d\phi^d}{d\tilde{t}}\right)
 +\frac{d}{d\tilde{t}}\left(\tilde{\nabla}_b\left(S_{ac}\frac{d\phi^c}{d\tilde{t}}\right)\right)\right],
\end{align}
which, on solving iteratively, at leading order gives
\begin{equation}\label{eq:4}
 \tilde{\nabla}_b\left(S_{ac}\frac{ d\phi^c}{d\tilde{t}}\right)\simeq-\frac{1}{3\tilde{H}}\left[3\tilde{H}_{,b}S_{ac}\frac{d\phi^c}{d\tilde{t}}+\tilde{V}_{;ab}+
 \tilde{R}_{adbc}\frac{d\phi^c}{d\tilde{t}}\frac{d\phi^d}{d\tilde{t}}\right].
\end{equation}
Note that we have made the assumption that the field velocity is not an
independent degree of freedom, such that
$d^2\phi^a/d\tilde{t}^2=d\phi^b/d\tilde{t}\tilde{\nabla}_b(d\phi^a/d\tilde{t})$.
Had we initially started with the slow-roll equations of motion
\eqref{eFsR}, then the curvature term in the above expression would not
have appeared, so let us confirm the conditions under which we are free
to neglect it.  We start by relating the Christoffel symbols in the
Jordan and Einstein frames by using the known form of $S_{ab}$, \eqref{albert}
\begin{align}\label{eq:10}
 S_{ad}\,^{(S)}\Gamma^d_{bc} = \frac{1}{2f}h_{ad}\,^{(h)}\Gamma^d_{bc} -
 \frac{1}{4f^2}\left(f_c h_{ab}+f_bh_{ac}-f_a h_{bc}\right)
 -\frac{3f_af_bf_c}{2f^3}+\frac{3}{2f^2}f_af_{bc}. 
\end{align}
As such, at leading order in the slow-roll approximation, the first two
terms on the right-hand side of \eqref{eq:4} are given as 
\begin{align}\nonumber
 3\tilde{H}_{,b}S_{ac}\frac{d\phi^c}{d\tilde{t}}+\tilde{V}_{;ab} &=
 \frac{W}{f}\left[-\frac{fW_aW_b}{8W^2}+\frac{fW_{ab}}{4W}-\frac{f\,^{(h)}\Gamma^c_{ab}W_c}{4W}\right.\\\label{eq:9}
&\quad\qquad\left.+\frac{1}{8}\left(\frac{f_aW_a}{W}+\frac{f_bW_a}{W}-h_{ab}\frac{f_cW^c}{W}\right)+\frac{3f_af_bf_cW^c}{4fW}-\frac{3f_{ab}f_cW^c}{4W}\right].
\end{align} 
Under the assumptions outlined in \eqref{eq:11}, we have that
$[...]\lesssim\mathcal{O}(\epsilon)$, where $[...]$ denotes all the
terms in the square brackets on the right-hand-side of \eqref{eq:9}.
Thus, if we would like to neglect the curvature term, we require that 
\begin{align}
 \frac{f}{W}\tilde{R}_{adbc}\frac{d\phi^c}{d\tilde{t}}\frac{d\phi^d}{d\tilde{t}}\ll\mathcal{O}(\epsilon)\quad\Rightarrow\quad f^2\tilde{R}_{adbc}\ll\mathcal{O}(1).
\end{align}
With $\tilde{R}_{abcd}$ given as
\begin{align}
 \tilde{R}_{abcd}=\left[S_{ae}\,^{(S)}\Gamma^e_{bd}\right]_c
 -\left[S_{ae}\,^{(S)}\Gamma^e_{bc}\right]_d - ^{(S)}\Gamma^e_{ac}S_{ef}\,^{(S)}\Gamma^f_{bd}+^{(S)}\Gamma^e_{ad}S_{ef}\,^{(S)}\Gamma^f_{bc}, 
\end{align}
and using \eqref{eq:10} to write
$S_{ad}\,^{(S)}\Gamma^d_{bc}=h_{ad}\,{}^{(h)}\Gamma^d_{bc}/(2f)+\Delta\Gamma_{bc|a}$,
where
\begin{equation}\label{eq:12}
 \Delta\Gamma_{ab|c} = -
 \frac{1}{4f^2}\left(f_c h_{ab}+f_bh_{ac}-f_a h_{bc}\right)
 -\frac{3f_af_bf_c}{2f^3}+\frac{3}{2f^2}f_af_{bc},
\end{equation}
we find that we can satisfy the above constraint if 
\begin{gather}
 fh_{ae}\left[^{(h)}\Gamma^e_{bd}\right]_c\ll1,\qquad 
 f_ch_{ae} \,^{(h)}\Gamma^e_{bd} \ll 1,\qquad
 f^2\left[\Delta\Gamma_{bd|a}\right]_c\ll 1,\qquad
 h_{ef}\,^{(h)}\Gamma^f_{ac}S^{eg}h_{gi}\,^{(h)}\Gamma^i_{bd} \ll 1,\\
 f\Delta\Gamma_{ac|e}S^{ef}h_{fg}\,^{(h)}\Gamma^g_{bd} \ll 1,\qquad
 f^2\Delta\Gamma_{ac|e}S^{ef}\Delta\Gamma_{bd|f} \ll 1.
\end{gather} 
Assuming that the conditions given in \eqref{eq:11} hold, that
$S^{ab}\sim 2fh^{ab}$ and $|h_{ab}|\sim\mathcal{O}(1)$, we can deduce
that the second and fourth of these conditions already hold.  The first
condition then gives an additional constraint on the Jordan frame
metric.  Finally, taking $X\ll 1$ to mean $X\lesssim\mathcal{O}(\epsilon)$,
the remaining three constraints are satisfied if  
\begin{equation}
 f^2\left[\Delta\Gamma_{bd|a}\right]_c\lesssim\mathcal{O}(\epsilon)\qquad\mbox{and}\qquad f^{3/2}\Delta\Gamma_{ab|c}\lesssim\mathcal{O}(\epsilon^{1/2}).
\end{equation}  
Referring to \eqref{eq:12}, the second of these will be satisfied if
$f_{ab}\lesssim\mathcal{O}(1)$, and differentiating
\eqref{eq:12} we find the requirements $f_{ab}\lesssim\mathcal{O}(\epsilon)$
and $\sqrt{f}f_{abc}\lesssim\mathcal{O}(\epsilon^{1/2})$.  In summary, we
are able to neglect the field-space curvature terms if, in addition to
the conditions outlined in \eqref{eq:11}, the conditions 
\begin{equation}
 f_{ab}\lesssim\mathcal{O}(\epsilon),\qquad
  \sqrt{f}f_{abc}\lesssim\mathcal{O}(\epsilon^{1/2})\qquad\mbox{and}\qquad
  \left[^{(h)}\Gamma^a_{bc}\right]_d \lesssim \mathcal{O}(\epsilon)
\end{equation} 
are satisfied.  Note that if we take $h_{ab}=\delta_{ab}$ in the
analytically soluble cases, the last of these will automatically be
satisfied.  Also note that as the first and second derivatives of $W$
and $f$ are constrained, this will also put constraints on the
derivatives of $V$.  Finally, we %comment
emphasise again that the above conditions are
sufficient rather than necessary conditions, as cancellation between
terms may relax the conditions, especially in the contraction of
field-space quantities.

\section{Non-minimally coupled spectator field model}\label{nmcsapp}

In this appendix we give additional details regarding the non-minimally coupled
spectator field model considered in the main text.  

In the Einstein frame, using
\begin{equation}
 \delta\tilde{\omega}
  =\left.\frac{\partial\tilde{\omega}}{\partial\phi}\right|_\diamond\delta\phi_{\tilde{\omega}}+\left.\frac{\partial\tilde{\omega}}{\partial\chi}\right|_\diamond\delta\chi_{\tilde{\omega}}=0\qquad\mbox{and}\qquad\frac{d\phi}{g^{(\phi)}}=\frac{d\chi}{g^{(\chi)}}
\end{equation} 
we find
\begin{equation}
\frac{\partial\phi_{\tilde{\omega}}}{\partial \phi_\ast} = 0, \qquad
\frac{\partial\phi_{\tilde{\omega}}}{\partial \chi_\ast} = 0,
\qquad\frac{\partial\chi_{\tilde{\omega}}}{\partial \phi_\ast}= 0
\qquad\mbox{and} \qquad \frac{\partial\chi_{\tilde{\omega}%actp
}}{\partial
\chi_\ast}= 1,
\end{equation} 
which leads to
\begin{subequations} %13:33equations
 \label{allequations}
\begin{align}
\tilde{N}_\phi &=
\frac{1}{g^{(\phi)}_\ast}\frac{(g^{(\phi)}_\diamond)^2}{2\epsilon_\diamond}=\frac{1}{g^{(\phi)}_\ast}\frac{1}{2f_\diamond},\\
\tilde{N}_\chi &= -\frac{1}{2}\int^\ast_\diamond \frac{f_{\chi}}{f^2}
\frac{1}{g^{(\phi)}}d\phi,\\ \tilde{N}_{\phi\phi} &=
\frac{1}{2f}\left(1-\frac{\eta^{(\phi)}_\ast}{2\epsilon_\ast}\right),\\
\tilde{N}_{\chi\chi}
&=%-\frac{1}{2g_\diamond^{(\phi)}}\frac{f_\chi^2}{f^3}
\frac{1}{2\epsilon_\diamond}\frac{W_{\chi\chi}}{W}-\frac{1}{2}\int^\ast_\diamond\left(\frac{f_{\chi\chi}}{f^2}-\frac{2(f_{\chi})^2}{f^3}\right)\frac{1}{g^{(\phi)}}d\phi,\\
\tilde{N}_{\phi\chi} &= -\frac{f_\chi}{2f^2g^{(\phi)}_\ast}.
\end{align}
\end{subequations}
Similarly, in the Jordan frame we have\footnote{Where the difference in comparison to the Einstein frame result is the
replacement $\tilde{\omega}\rightarrow\omega$.}
\begin{equation}
\frac{\partial\phi_\omega}{\partial \phi_\ast} = 0, \qquad
\frac{\partial\phi_\omega}{\partial \chi_\ast} =
-\frac{f_\chi}{f}\frac{1}{g^{(\phi)}_\omega},
\qquad\frac{\partial\chi_\omega}{\partial \phi_\ast}= 0 \qquad\mbox{and}
\qquad \frac{\partial\chi_\omega}{\partial \chi_\ast}= 1,
\end{equation} 
giving us
\begin{subequations} %13:35equations
\begin{align}
N_\phi &=\frac{1}{g^{(\phi)}_\ast}\frac{1}{2f_\diamond}=\tilde{N}_\phi,
\\ N_\chi &= \frac{f_{\chi}}{2f_\diamond\epsilon_\diamond}-\frac{1}{2}\int^\ast_\diamond
\frac{f_{\chi}}{f^2} \frac{1}{g^{(\phi)}}d\phi =
\tilde{N}_\chi + \frac{f_{\chi}}{2f_\diamond\epsilon_\diamond},\\ N_{\phi\phi} &=
\frac{1}{2f}\left(1-\frac{\eta^{(\phi)}_\ast}{2\epsilon_\ast}\right)=
\tilde{N}_{\phi\phi},\\ N_{\chi\chi} &=
%\frac{1}{\epsilon_\diamond}\left(\frac{f_{\chi\chi}}{2f} -
%\frac{(f_{\chi})^2}{f^2}\left(2-\frac{\eta^{(\phi)}_\diamond}{2\epsilon_\diamond}\right)\right)-\frac{1}{2}\int^\ast_\diamond\left(\frac{f_{\chi\chi}}{f^2}-\frac{2(f_{\chi})^2}{f^3}\right)\frac{1}{g^{(\phi)}}d\phi
%\\ &= 
\tilde{N}_{\chi\chi} +
%\frac{1}{2g_\diamond^{(\phi)}}\frac{f_\chi^2}{f^3} +
\frac{1}{2\epsilon_\diamond}\left(\frac{f_{\chi\chi}}{f} 
-2\frac{(f_\chi)^2}{f\epsilon_\diamond}\frac{W_{\chi\chi}}{W}-
\frac{(f_{\chi})^2}{f^2}\left(5-\frac{\eta^{(\phi)}_\diamond}{\epsilon_\diamond}\right)\right),\\
N_{\phi\chi} &= -\frac{f_\chi}{2f^2g^{(\phi)}_\ast} =
\tilde{N}_{\phi\chi}.
\end{align}
\end{subequations}

In the case of the specific example considered in the text, where
$V=V^{(\chi)}(\chi)V^{(\phi)}(\phi)$ and $V^{(\phi)}(\phi)=m^2\phi^{2p}$,
we have
\begin{equation}
 N =
  \frac{1}{8fp}\left[\phi_\ast^2-\phi_\diamond^2\right]\quad\mbox{and}\quad
  \epsilon_\diamond = \frac{4p^2f}{\phi_\diamond^2}.
\end{equation}
As such, taking the end of inflation to correspond to $\epsilon_\diamond
= 1$ and defining $N_\ast$ as the number of
e-folds before the end of inflation that the scale under consideration
left the horizon, we have
\begin{subequations} %18:00equations
\begin{align}
 \tilde{N}_\phi &= \left(\frac{2N_\ast+p}{4pf}\right)^{1/2},\\
\tilde{N}_\chi &= -\frac{f_\chi}{f}N,\\
\tilde{N}_{\phi\phi} &= \frac{1}{4fp},\\
\tilde{N}_{\phi\chi} &= -\frac{f_\chi}{f}\left(\frac{2N_\ast+p}{4pf}\right)^{1/2},\\
\tilde{N}_{\chi\chi} &= \frac{W_{\chi\chi}}{2pW}\left(2(N_\ast-N)+p\right)-\frac{N}{f}\left(f_{\chi\chi}-\frac{2f_\chi^2}{f}\right),
\end{align}  
\end{subequations}
and 
\begin{align}
 N_\phi&=\tilde{N}_\phi,\qquad N_{\phi\phi} = \tilde{N}_{\phi\phi}, \qquad
 N_{\phi\chi}=\tilde{N}_{\phi\chi},\\
 N_\chi &= \tilde{N}_\chi +
 \frac{f_\chi(2(N_\ast-N)+p)}{2fp},\\
 N_{\chi\chi} &= \tilde{N}_{\chi\chi}+\frac{2(N_\ast-N)+p}{2pf}\left(f_{\chi\chi}-\frac{f_\chi^2(3p+1)}{pf}-\frac{2(N_\ast-N)+p}{p}\frac{2f_\chi^2W_{\chi\chi}}{W}\right).
\end{align}

%{\bf
%Assuming that the fluctuations of the spectator field 
%are the dominant source of the curvature perturbations 
%at the end of inflation
%where the difference of spectral quantities
%between the conformal frames
%is negligible,
%and setting $f=1/2$,
%we find
%\begin{eqnarray}
%\tilde{n}_s-1\approx -\frac{2}{f_\chi N},\quad
%\tilde{r}\approx \frac{2(1+f_\chi^2)}{f_\chi^2 N^2}.
%\end{eqnarray}
%Thus we have to choose a positive $f_\chi$
%to get a red-tilted spectrum for the curvature perturbations.
%Since
%\begin{eqnarray}
%\tilde{r}\approx \frac{1+6f_\chi^2}{f_\chi N}(1-\tilde{n}_s),
%\end{eqnarray}
%for $f_\chi<1$,
%%as $f_\chi$ increases $\tilde{r}$ decreases.
%A larger $N$ also makes $\tilde{r}$ smaller.
%
%As $f_\chi \to 0$, the above ratio appears to be divergent, but 
%in this case the above relations are not valid 
%because the inflaton fluctuation
%becomes the dominant source of the curvatre perturbations.
%In this limit, it is easy to confirm that the results for
%a single-field inflation model are recovered: 
%\begin{eqnarray}
%{\tilde n}_s-1\approx-6\epsilon_\ast +2\eta_\ast^{(\phi)},
%\quad
%{\tilde r}\approx 8\big( g_\ast^{(\phi)}\big)^2.
%\end{eqnarray}
%Thus, for a given spectral index,
%adding a non-minimal coupling to the spectator field with $f_\chi>0$
%contributes to reduce the tensor-to-scalar ratio. 
%%}%

\section{Non-minimally coupled multi-brid inflation}\label{AppNMM}

In this appendix we give details of the non-minimally coupled extension
of the multi-brid model of inflation \cite{multibrid}.

Taking our potential and non-minimal coupling to be of the form given in
(\ref{mb_pot}), the slow-roll parameters (\ref{sR1})-(\ref{sR3}) are then given explicitly as
\begin{eqnarray}\label{eq:14}
\epsilon &=&
\sqrt{f_0}\exp\left[\sum_a\frac{z_a}{2}\phi^a\right]\sum_b(m_b-z_b)^2,\\\label{sr2}
\eta^{(a)} & =&
2\sqrt{f_0}\exp\left[\sum_a\frac{z_a}{2}\phi^a\right]\sum_bm_b(m_b-z_b),\\\label{sr3}
\delta
&=&\frac{1}{6}\sqrt{f_0}\exp\left[\sum_a\frac{z_a}{2}\phi^a\right]\sum_bz_b(m_b-z_b).
\end{eqnarray}
Referring to Eq. (\ref{analCon}),
we also find 
\begin{eqnarray}
\quad g^{(a)} := m_a-z_a\quad\mbox{and} \quad 
F =\frac{1}{2f}
=
\frac{1}{2\sqrt{f_0}\exp\left[\sum_a\frac{z_a}{2}\phi^a\right]},
%\\
\end{eqnarray}
which gives us
\begin{align}
 &\quad q^a =
\exp\left[\frac{\phi^a}{m_a - z_a}\right].
%\\ &\quad \ln q = \ln n^{1/2}
%+\frac{\phi^a}{m_a-z_a}\quad\mbox{for any a},
\end{align}
The slow-roll equations of motion (\ref{eom_sR}) can then be expressed as 
\begin{align} 
%\Rightarrow& \quad 
\frac{d\ln q^a}{dN} = \frac{d\ln q}{dN} =
-2\sqrt{f_0}\exp\left[\sum_a\frac{z_a}{2}\phi^a\right]=-2\sqrt{f_0}q^{\sum_az_a(m_a-z_a)/2}\prod_a
 (n^a)^{z_a(m_a-z_a)},
\end{align}
which we can integrate to find
\begin{equation}\label{eq:17}
N =
\frac{1}{\sqrt{f_0}\sum_az_a(m_a-z_a)}\left[\exp\left[-\sum_b\frac{z_b}{2}\phi_\diamond^b\right]
-\exp\left[-\sum_b\frac{z_b}{2}\phi_\ast^b\right]\right].
\end{equation}
The condition for the end of inflation is given by the tachyonic
instability of the waterfall field as discussed in section \ref{nmmbsec}.  In
order to relate the field values at the end of inflation to those at the
beginning, we note that as the $n^a$s are constant, we have the invariant
$\ln (q^1/q^2)$, which allows us to write
\begin{equation}
\frac{\phi^1_\ast}{m_1-z_1} - \frac{\phi^2_\ast}{m_2-z_2} =
\frac{\sigma\cos\gamma}{w_1(m_1-z_1)} -
\frac{\sigma\sin\gamma}{w_2(m_2-z_2)}.
\end{equation}
Perturbing this expression to second order we find
\begin{align}\nonumber
\frac{\delta\phi^1_\ast}{m_1-z_1} - \frac{\delta\phi^2_\ast}{m_2-z_2} &=
-\left(\frac{\sigma\sin\gamma}{w_1(m_1-z_1)} +
\frac{\sigma\cos\gamma}{w_2(m_2-z_2)}\right)(\delta\gamma_{(1)} +
\delta\gamma_{(2)})\\ &\quad-\frac{\sigma\cos\gamma}{2w_1(m_1-z_1)}\delta\gamma_{(1)}^2
+ \frac{\sigma\sin\gamma}{2w_2(m_2-z_2)}\delta\gamma_{(1)}^2,
\end{align}
and by solving order by order we are able to find expressions for
$\delta\gamma_{(1)}$ and $\delta\gamma_{(2)}$ in terms of
$\delta\phi^1_\ast$ and $\delta\phi^2_\ast$.  
Finally, expanding $\delta
N^\sigma$ up to second order as
\begin{equation}
\delta N^\sigma = \sum_a N^\sigma_a\delta\phi_\ast^a
+ N^\sigma_\gamma(\delta\gamma_{(1)} +
\delta\gamma_{(2)})+\sum_{a,b}\frac{1}{2}N^\sigma_{ab}\delta\phi^a_\ast\delta\phi^b_\ast
+
\sum_a N^\sigma_{a\gamma}\delta\phi^a_\ast\delta\gamma_{(1)}+\frac{1}{2}N^\sigma_{\gamma\gamma}\delta\gamma_{(1)}^2,
\end{equation}
where
\begin{subequations}
\begin{align}
\delta\gamma_{(1)} &=
-\frac{w_1w_2}{\sigma}\frac{(m_2-z_2)\delta\phi^1_\ast-(m_1-z_1)\delta\phi^2_\ast}{(m_2-z_2)w_2\sin\gamma
+ (m_1-z_1)w_1\cos\gamma},\\ \delta\gamma_{(2)} &=
\frac{\delta\gamma_{(1)}^2}{2}\frac{(m_1-z_1)w_1\sin\gamma -
(m_2-z_2)w_2\cos\gamma}{(m_2-z_2)w_2\sin\gamma +
(m_1-z_1)w_1\cos\gamma},\\\label{mulhyb}
N^\sigma_1&=
\frac{z_1}{2f_\ast(z_1(m_1-z_1)+z_2(m_2-z_2))},\\
N^\sigma_2&=\frac{z_2}{2f_\ast(z_1(m_1-z_1)+z_2(m_2-z_2))},\\
N^\sigma_\gamma&=\frac{\frac{z_1\sigma}{w_1}\sin\gamma-\frac{z_2\sigma}{w_2}\cos\gamma}{2f_\diamond(z_1(m_1-z_1)+z_2(m_2-z_2))},\\ 
N^\sigma_{11}&=-\frac{z^2_1}{4f_\ast(z_1(m_1-z_1)+z_2(m_2-z_2))},\\
N^\sigma_{22}&=-\frac{z^2_2}{4f_\ast(z_1(m_1-z_1)+z_2(m_2-z_2))},\\
N^\sigma_{12}&=-\frac{z_1z_2}{4f_\ast(z_1(m_1-z_1)+z_2(m_2-z_2))},\\\label{mulhybend}
N^\sigma_{\gamma\gamma}&=\frac{\frac{z_1\sigma}{w_1}\cos\gamma+\frac{z_2\sigma}{w_2}\sin\gamma+\frac{1}{2}\left(\frac{z_1\sigma}{w_1}\sin\gamma-\frac{z_2\sigma}{w_2}\cos\gamma\right)^2}{2f_\diamond(z_1(m_1-z_1)+z_2(m_2-z_2))},
\end{align}
\end{subequations}
we arrive at the final expression \eqref{eq:13}.

In the main text we discuss two contributions to the final $\delta N$:
that arising from evolution up to the end-of-inflation surface and that arising from the evolution between the
end-of-inflation surface and a constant energy surface sometime
during the radiation domination era.  As discussed in \cite{multibrid},
it is also possible to consider the $\delta N$ arising from the
evolution up to some constant energy surface during
inflation.  In our case, the definition of the constant energy surface depends on the frame.  In the Jordan
frame we have $\omega = V/2f = \rm{const.}$.  Expliclty
\begin{equation}
\frac{V_0}{2\sqrt{f_0}}\exp\left[\sum_a\left(m_a-\frac{z_a}{2}\right)\phi^a_\diamond\right]
= \rm{const.}.
\end{equation}
Perturbing this quantity gives us
\begin{equation}
\left(m_1-\frac{z_1}{2}\right)\delta\phi^1_\rho +
\left(m_2-\frac{z_2}{2}\right)\delta\phi^2_\rho=0,
\end{equation}
where we use the $\rho$ subscript to denote the field perturbations on a surface of constant
energy during inflation.  If we then exploit the fact that $\ln(q_1/q_2) = \rm{const.}$, which
gives us
\begin{equation}
\frac{\phi^1_\ast}{m_1-z_1} - \frac{\phi^2_\ast}{m_2-z_2} =
\frac{\phi^1_\diamond}{m_1-z_1} - \frac{\phi^2_\diamond}{m_2-z_2},
\end{equation}
we are able to relate the perturbations on the constant energy
hypersurface with those on the initial flat hypersurface.  For example,
we have
\begin{equation}
\delta\phi^1_\rho = \frac{(m_2-z_2/2)(m_2-z_2)\delta\phi^1_\ast -
(m_2-z_2/2)(m_1-z_1)\delta\phi^2_\ast}{(m_2-z_2/2)(m_2-z_2)+(m_1-z_1/2)(m_1-z_1)}.
\end{equation}
As such, we find
\begin{subequations} %14:14equations
\begin{align}
%\frac{\partial N_\rho}{\partial \phi^1_\ast} 
N^\rho_{1}&= \frac{1}{\sum_az_a(z_a -
m_a)}\left\{-\frac{z_1}{2f_\ast} +
\frac{1}{2f_\diamond}\frac{(z_1m_2-z_2m_1)(m_2-z_2)}{(m_2-z_2/2)(m_2-z_2) +
(m_1-z_1/2)(m_1-z_1)}\right\},\\ 
%\frac{\partial N_\rho}{\partial\phi^1_\ast} 
N^\rho_{2}&= \frac{1}{\sum_az_a(z_a -
m_a)}\left\{-\frac{z_2}{2f_\ast} -
\frac{1}{2f_\diamond}\frac{(z_1m_2-z_2m_1)(m_1-z_1)}{(m_2-z_2/2)(m_2-z_2) +
(m_1-z_1/2)(m_1-z_1)}\right\},\\ 
%\frac{\partial^2 N_\rho}{\partial(\phi^1_\ast)^2} 
N^\rho_{11}&= \frac{1}{\sum_az_a(z_a -
m_a)}\left\{\frac{z^2_1}{4f_\ast} -
\frac{1}{4f_\diamond}\left[\frac{(z_1m_2-z_2m_1)(m_2-z_2)}{(m_2-z_2/2)(m_2-z_2)
+ (m_1-z_1/2)(m_1-z_1)}\right]^2\right\},\\ 
%\frac{\partial^2N_\rho}{\partial (\phi^2_\ast)^2} 
N^\rho_{22}&= \frac{1}{\sum_az_a(z_a -
m_a)}\left\{\frac{z^2_2}{4f_\ast} -
\frac{1}{4f_\diamond}\left[\frac{(z_1m_2-z_2m_1)(m_1-z_1)}{(m_2-z_2/2)(m_2-z_2)
+ (m_1-z_1/2)(m_1-z_1)}\right]^2\right\},\\ 
%\frac{\partial^2N_\rho}{\partial \phi^1_\ast\phi^2_\ast} 
N^\rho_{12}&= \frac{1}{\sum_az_a(z_a -
m_a)}\left\{\frac{z_1z_2}{4f_\ast} +
\frac{1}{4f_\diamond}\frac{(z_1m_2-z_2m_1)^2(m_1-z_1)(m_2-z_2)}{\left[(m_2-z_2/2)(m_2-z_2)
+ (m_1-z_1/2)(m_1-z_1)\right]^2}\right\}.
\end{align}
\end{subequations}
In the Einstein frame we have $\tilde{\omega} = W/4 = {\rm const.}$, or
explicitly
\begin{equation}
\frac{V_0}{4f_0}\exp\left[\sum_a(m_a - z_a)\phi^a_\diamond\right] = {\rm const.}.
\end{equation}
We consequently find
\begin{subequations} %14:14equations
\begin{align}
%\frac{\partial N_\rho}{\partial \phi^1_\ast} 
N^{\tilde{\rho}}_{1}&= \frac{1}{\sum_az_a(z_a -
m_a)}\left\{-\frac{z_1}{2f_\ast} +
\frac{1}{2f_\diamond}\frac{(z_1m_2-z_2m_1)(m_2-z_2)}{(m_2-z_2)^2 +
(m_1-z_1)^2}\right\},\\ 
%\frac{\partial N_\rho}{\partial \phi^1_\ast} 
N^{\tilde{\rho}}_2&=\frac{1}{\sum_az_a(z_a - m_a)}\left\{-\frac{z_2}{2f_\ast} -
\frac{1}{2f_\diamond}\frac{(z_1m_2-z_2m_1)(m_1-z_1)}{(m_2-z_2)^2 +
(m_1-z_1)^2}\right\},\\ 
%\frac{\partial^2 N_\rho}{\partial (\phi^1_\ast)^2}
N^{\tilde{\rho}}_{11}&= \frac{1}{\sum_az_a(z_a - m_a)}\left\{\frac{z^2_1}{4f_\ast} -
\frac{1}{4f_\diamond}\left[\frac{(z_1m_2-z_2m_1)(m_2-z_2)}{(m_2-z_2)^2 +
(m_1-z_1)^2}\right]^2\right\},\\ 
%\frac{\partial^2 N_\rho}{\partial(\phi^2_\ast)^2} 
N^{\tilde{\rho}}_{22}&= \frac{1}{\sum_az_a(z_a -
m_a)}\left\{\frac{z^2_2}{4f_\ast} -
\frac{1}{4f_\diamond}\left[\frac{(z_1m_2-z_2m_1)(m_1-z_1)}{(m_2-z_2)^2 +
(m_1-z_1)^2}\right]^2\right\},\\ 
%\frac{\partial^2 N_\rho}{\partial\phi^1_\ast\phi^2_\ast} 
N^{\tilde{\rho}}_{12}&= \frac{1}{\sum_az_a(z_a -
m_a)}\left\{\frac{z_1z_2}{4f_\ast} +
\frac{1}{4f_\diamond}\frac{(z_1m_2-z_2m_1)^2(m_1-z_1)(m_2-z_2)}{\left[(m_2-z_2)^2
+ (m_1-z_1)^2\right]^2}\right\}.
\end{align}
\end{subequations}
These results are in contrast with the minimally coupled case, where the
second derivatives of $N$ are all exactly zero and there is therefore no
non-gaussianity induced during inflation.  However, even in this
non-minimally coupled case we see that if
$m_a,z_a\sim\mathcal{O}(\epsilon^{1/2})$ (which is required for the
slow-roll conditions to hold) then $N_{ab}\sim\mathcal{O}(1)$, meaning
that the associated $f_{NL}$ parameter will be slow-roll suppressed.

\end{document}